\documentclass[trackchanges]{aastex7} 
\usepackage[ruled,vlined]{algorithm2e}
\usepackage{amsmath}
\usepackage{CJKutf8}
\usepackage{csquotes}
\usepackage{siunitx}
\usepackage{tabularx, makecell, booktabs}
\usepackage{graphicx}
\usepackage{subfig}
\hypersetup{
    colorlinks=true,
    linkcolor=blue,
    filecolor=magenta,      
    urlcolor=cyan,
}

\begin{document}
\title{Projection Is All You Need: Interpreting Polarization Measurements in the Orion Clouds with Sub-Alfv\'{e}nic MHD Simulations}

\author[orcid=0009-0006-9160-1021, sname=Quan]{Feiyu Quan \texorpdfstring{\protect\begin{CJK*}{UTF8}{gbsn}(全飞宇)\protect\end{CJK*}}{}}
\affiliation{Department of Physics, The Chinese University of Hong Kong, Shatin, Hong Kong, China}
\email{feiyu.quan@link.cuhk.edu.hk} 

\author[orcid=0009-0000-2782-2862, sname=Xu]{Yitao Xu \texorpdfstring{\protect\begin{CJK*}{UTF8}{gbsn}(徐义涛)\protect\end{CJK*}}{}}
\affiliation{Department of Statistics and Data Science, The Chinese University of Hong Kong, Shatin, Hong Kong, China}
\email{yitaoxu@link.cuhk.edu.hk}

\author[orcid=0000-0002-5093-5088, sname=Qiu]{Keping Qiu \texorpdfstring{\protect\begin{CJK*}{UTF8}{gbsn}(邱科平)\protect\end{CJK*}}{}}
\affiliation{School of Astronomy and Space Science, Nanjing University, 163 Xianlin Avenue, Nanjing 210023, Jiangsu, People's Republic of China}
\affiliation{Key Laboratory of Modern Astronomy and Astrophysics (Nanjing University), Ministry of Education, Nanjing 210023, Jiangsu, People's
Republic of China}
\email{kpqiu@nju.edu.cn}

\author[orcid=0000-0002-2744-9030, sname=Fan]{Xiaodan Fan \texorpdfstring{\protect\begin{CJK*}{UTF8}{gbsn}(樊晓丹)\protect\end{CJK*}}{}}
\affiliation{Department of Statistics and Data Science, The Chinese University of Hong Kong, Shatin, Hong Kong, China}
\email[show]{xfan@cuhk.edu.hk}

\author[orcid=0000-0003-2641-9240, sname=Li]{Hua-bai Li \texorpdfstring{\protect\begin{CJK*}{UTF8}{bsmi}(李華白)\protect\end{CJK*}}{}}
\affiliation{Department of Physics, The Chinese University of Hong Kong, Shatin, Hong Kong, China}
\email[show]{hbli@cuhk.edu.hk}



\begin{abstract}
    Dust polarization observations are widely used to diagnose the relative importance of magnetic fields and turbulence in star forming molecular clouds, often through summary statistics such as the mean polarization direction $\mu$ and dispersion $\sigma$. Recent multi-scale polarization observations of the Orion Integral-Shaped Filament (ISF) reveal substantial diversity in polarization morphology among its dense cores, raising questions about the underlying Alfvénic nature of the cloud. In this work, we develop a statistical framework to compare polarization-based summary statistics from observations with those derived from projected three dimensional MHD simulations, explicitly accounting for projection effects. Using globally sub-Alfvénic simulations that naturally produce slightly super-Alfvénic dense cores, we show that modest deviations of core-scale magnetic fields from the parent cloud field, when combined with projection, can generate a wide range of plane-of-sky polarization dispersions. Applying hypothesis testing, we find that the observed $(\mu, \sigma)$ values in the Orion ISF are statistically consistent with sub-Alfvénic cloud models over a broad range of viewing angles. This broad degeneracy implies that $\mu$ and $\sigma$ alone cannot provide precise information about the three-dimensional magnetic-field distribution, and hence the Alfvén Mach number, of an individual cloud. While the observations can provide evidence against certain projection geometries, we demonstrate that polarization statistics based solely on $(\mu, \sigma)$ are insufficient to provide evidence against sub-Alfvénic cloud models. Our results highlight the necessity of explicitly incorporating projection effects when interpreting polarization observations of molecular clouds.


\end{abstract}



\setcounter{footnote}{0}

\section{Introduction} \label{sec:introduction}

\setcounter{footnote}{0}

Dust polarization observations are widely used to diagnose the relative importance of magnetic fields and turbulence in star forming molecular clouds \citep{1987ARA&A..25...23S, 2007ARA&A..45..565M, 2021Galax...9...41L}. A common observational practice is to characterize plane of sky magnetic field morphology using summary statistics such as the angular dispersion, $\sigma$, and the mean offset, $\mu$, of polarization angles relative to the large-scale mean field. In many studies, a larger polarization dispersion is interpreted as reflecting a higher level of turbulent perturbation relative to the magnetic field, whereas a small dispersion is often taken to imply a magnetically dominated (sub-Alfvénic) environment (e.g., \citealt{Hildebrand_2009, 2016A&A...586A.138P}).

It is important to note that such diagnoses are inherently based on projected quantities. Even classical methods such as the Davis–Chandrasekhar–Fermi (DCF) technique \citep{1951PhRv...81..890D, 1953ApJ...118..113C}, which relies on the observed plane of sky dispersion, are explicitly formulated to estimate the plane of sky magnetic field strength, not the total three dimensional field. As such, DCF-type reasoning already implicitly acknowledges that projection plays a central role in interpreting polarization data. Nevertheless, in practice, projection effects are often treated implicitly or assumed to be sub-dominant when polarization dispersion is used as a comparative diagnostic of Alfvén Mach number across regions.

Recent polarization observations of the Orion Integral-Shaped Filament (ISF) offer a particularly instructive illustration. Using JCMT and CARMA data from the BISTRO \citep{Wu_2024} and TADPOL surveys \citep{2014ApJS..213...13H}, \cite{Wu_2024} quantified the distributions of magnetic field orientation offsets across multiple spatial scales in OMC-1, OMC-2, and OMC-3. These regions, while belonging to the same molecular cloud, exhibit markedly different values of $\mu$ and $\sigma$. Such diversity naturally motivates interpretations in terms of varying magnetic or turbulent regimes and reflects a very general observational approach: using $(\mu, \sigma)$ as a proxy for the Alfvénic nature of the underlying cloud.

However, numerical studies of magnetized molecular clouds have shown that this interpretation can be incomplete if projection effects are not treated explicitly. In particular, simulations by \cite{2019ApJ...871...98Z}, \cite{Cao_2023}, and \cite{Yuan_2025} demonstrate that even in globally sub-Alfvénic clouds, dense cores can develop slightly locally super-Alfvénic turbulence (with Alfvén Mach number $\mathcal{M_A} \sim 2 -3$). \footnote{In this work, the Alfvén Mach number is defined as
     $$ \langle \mathcal{M}_A \rangle = \displaystyle \frac{\sigma_v}{\langle v_A \rangle}, $$
where $\sigma_v$ is the 3D velocity dispersion of a specified domain, $v_A$ is the local Alfv\'{e}n velocity, and $\langle \cdot \rangle$ denotes averaging over the domain.} Crucially, this level of turbulence is not sufficient to tangle the magnetic field and generate a large intrinsic three dimensional dispersion of field directions. Instead, it preferentially deviates the mean magnetic field of individual cores from that of their parent cloud. As a result, different cores within the same sub-Alfvénic cloud naturally possess different three dimensional orientations of their mean magnetic fields relative to the line of sight.

This distinction between tangling and deviation is critical. While the intrinsic 3D magnetic field within each core remains relatively ordered, the diversity of core scale mean field orientations leads to a wide range of observed plane of sky polarization dispersions once projected. Cores whose mean fields are oriented closer to the line of sight appear more disordered in polarization, whereas those viewed closer to perpendicular appear more ordered -- even though all originate from the same underlying sub-Alfvénic cloud. Section \ref{sec:projections} is dedicated to illustrating this subtle but important effect using explicit visualizations from simulations.

The goal of this work is therefore not to advocate for sub-Alfvénic cloud models, but to assess the diagnostic power of $(\mu, \sigma)$ itself. Specifically, we ask whether the values of $\mu$ and $\sigma$ measured in the Orion ISF are sufficient, on their own, to provide evidence against sub-Alfvénic cloud models once projection effects are properly taken into account. By combining sub-Alfvénic MHD simulations with systematic projection and a statistical comparison framework, we show that the observed $(\mu, \sigma)$ values in Orion are statistically consistent with sub-Alfvénic clouds over a broad range of viewing angles. Consequently, while polarization dispersion remains a useful descriptive observable, $(\mu, \sigma)$ alone does not provide a sufficient basis for rejecting sub-Alfvénic cloud models in the Orion ISF.

The remainder of this paper is organized as follows. In Section \ref{sec:projections}, we use visualizations from sub-Alfvénic MHD simulations to illustrate how projection effects, combined with modest core scale field deviations, can produce a wide diversity of observed plane of sky polarization morphologies without requiring intrinsically tangled magnetic fields. Section \ref{sec:simulation} describes the numerical setup of the sub-Alfvénic MHD simulations employed in this work. In Section \ref{sec:methods}, we detail the methodology for projecting the three dimensional simulation data into two dimensional polarization maps and for constructing summary statistics and probability distributions in the $(\mu, \sigma)$ space. Section \ref{sec:results} presents the statistical comparison between the Orion ISF observations and the projected simulations, including both maximum likelihood estimation and hypothesis testing. In Section \ref{sec:discussion}, we discuss the limitations of our current approach and its potential extension as a method for probing the three dimensional orientation of cloud scale magnetic fields. Finally, Section \ref{sec:conclusion} summarizes our main conclusions.

\begin{figure*}[ht]

\subfloat[Left: misaligned (between the red and the blue) magnetic field lines. Right: aligned magnetic field lines. Viewed along the $y$-axis of the domain, with the elevation angle = $0^\circ$. ]{%
  \includegraphics[clip,width=\columnwidth]{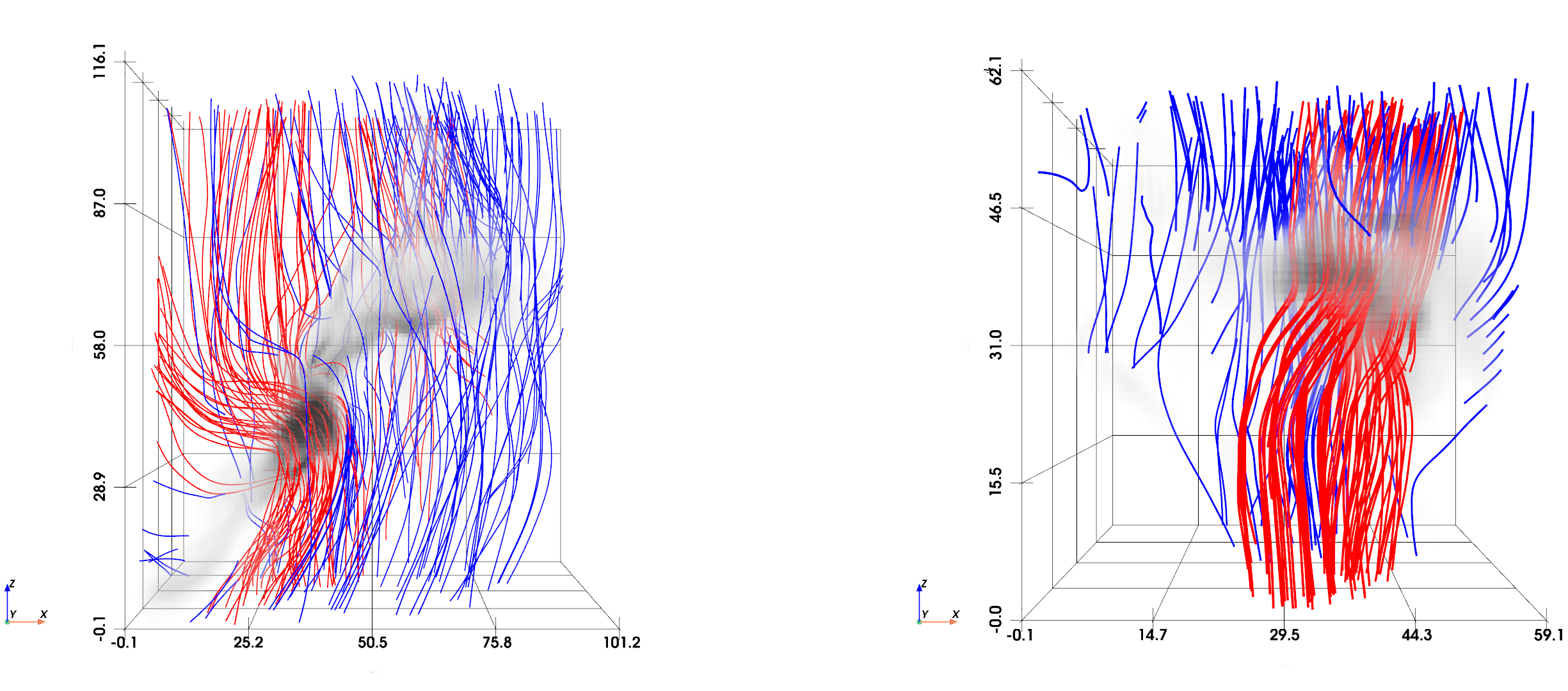}%
}

\subfloat[The same dense cores as in the top panel, but being viewed from a different perspective with elevation angle = $-20^\circ$ and azimuthal angle = $-145^\circ$. For the left one, the $90^\circ$ deviation of the red field lines around the densest region (as shown in the top panel) disappears and we can see more alignment between the red and blue field lines; while for the right one, the alignment between magnetic field lines stays largely the same as we switch the projection angle from top to bottom panel.]{%
  \includegraphics[clip,width=\columnwidth]{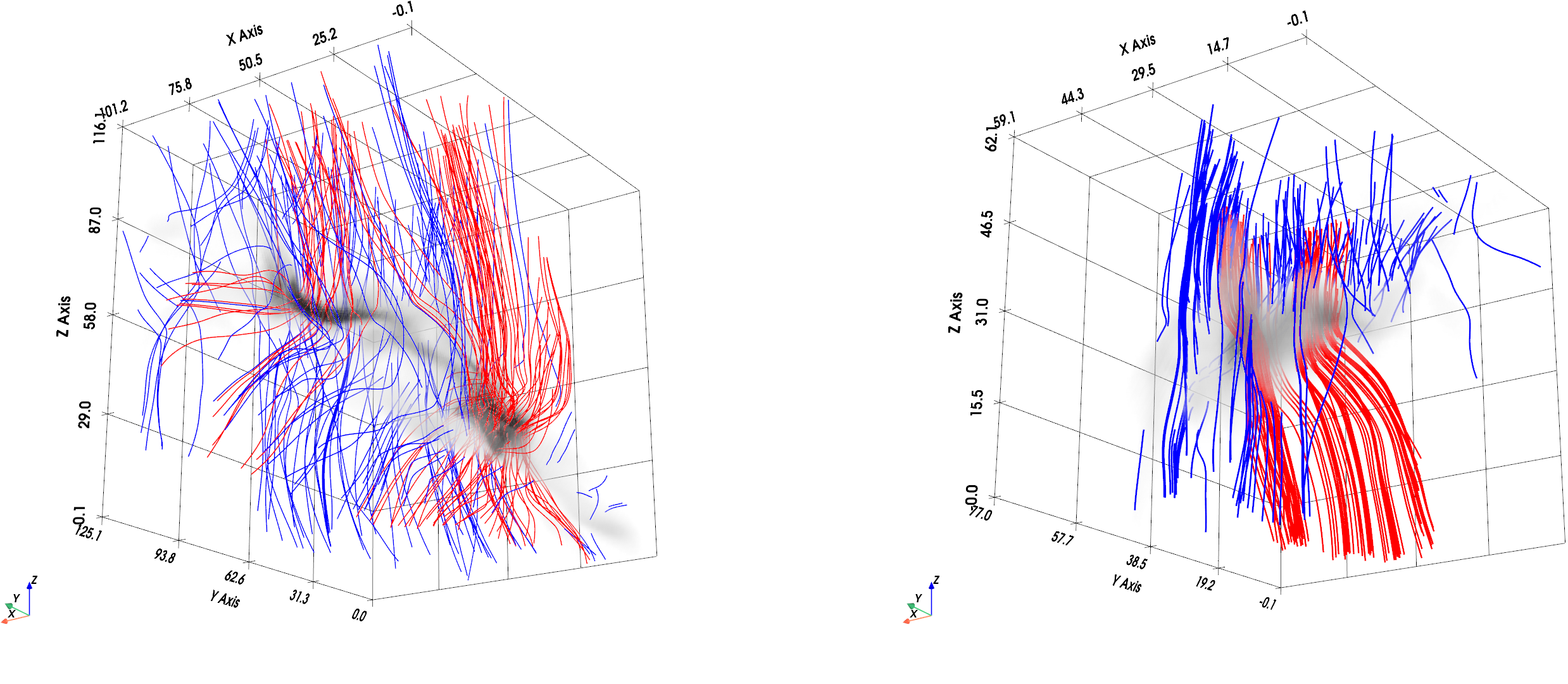}%
}

\caption{Two cases of selected dense cores and the magnetic field lines that pass through them from two different viewing angles, cut from two different sub-regions in one of our simulations towards the end of Phase II (the collapsing phase). \textbf{\textcolor{red}{Red field lines}} indicate that they pass through dense regions with volume density $> 5 \times 10^4 \ {\rm H_2}\ \mathrm{cm}^{-3}$, while the \textbf{\textcolor{blue}{blue field lines}} pass through diffuse gas with volume density $< 1 \times 10^4 \ {\rm H_2}\ \mathrm{cm}^{-3}$. The grey-scale color indicates different volume densities. The length of each side of the boxes is roughly 0.5 pc for the left one and 0.3 pc for the right one. An interactive version of this figure can be found at \href{https://www.phy.cuhk.edu.hk/sfg/Feiyu/clump2.html}{this https URL} and \href{https://www.phy.cuhk.edu.hk/sfg/Feiyu/clump4_new.html}{this https URL}. Visualization made using \texttt{PyVista} \citep{2019JOSS....4.1450S}.
\label{fig:dense_cores}}
\end{figure*}


\section{The Influence of Projection on Polarization Patterns} \label{sec:projections} 

Projection effect is a critical factor that affect the observed polarization maps from molecular clouds (e.g., \citealt{Basu_2000, 2020ApJ...899...28D, 2025A&A...696A..35T}). Prior to the rigorous statistical proof that the observed polarization maps in Orion ISF \citep{2014ApJS..213...13H, Wu_2024} do not provide enough evidence against a sub-Alfv\'{e}nic (strong-field) cloud model, we first present illustrative examples from simulations to highlight the diversity of possible magnetic field morphologies and alignments due to projections.

\subsection{Core-Cloud Field Alignment (\enquote{\texorpdfstring{$\mu$}{Lg}} in Table \ref{tab:parameter_definitions}) is Perspective- and Density-Dependent} \label{sec:2.1}

While the simulation details are provided in Section \ref{sec:simulation}, here we first illustrate the resulting \textit{B}-field morphologies using two sub-regions from the simulations in this work. For each subregion, Figure \ref{fig:dense_cores} shows the magnetic field from two different viewing angles (four panels in total).  Red field lines indicate that they pass through dense cores with volume density $> 5 \times 10^4 \ {\rm H_2}\ \mathrm{cm}^{-3}$, while the blue field lines pass through diffuse gas with volume density $< 1 \times 10^4 \ {\rm H_2}\ \mathrm{cm}^{-3}$. The grey-scale color indicates different volume densities. From the first perspective (top row), the two regions exhibit significantly different field alignments between high (red) and low (blue) densities --- one is well-aligned, while the other is nearly perpendicular. However, from the second perspective (bottom row), the field direction shows no significant change with density in either case.  This shows that we can observe diverse magnetic field morphologies from different dense cores within the same molecular cloud (just like OMC 1, 2 and 3 in Orion ISF) under our simulation setup, purely as a result of projection effects.

\begin{figure*}[ht]
\centering
\includegraphics[scale = 0.12]{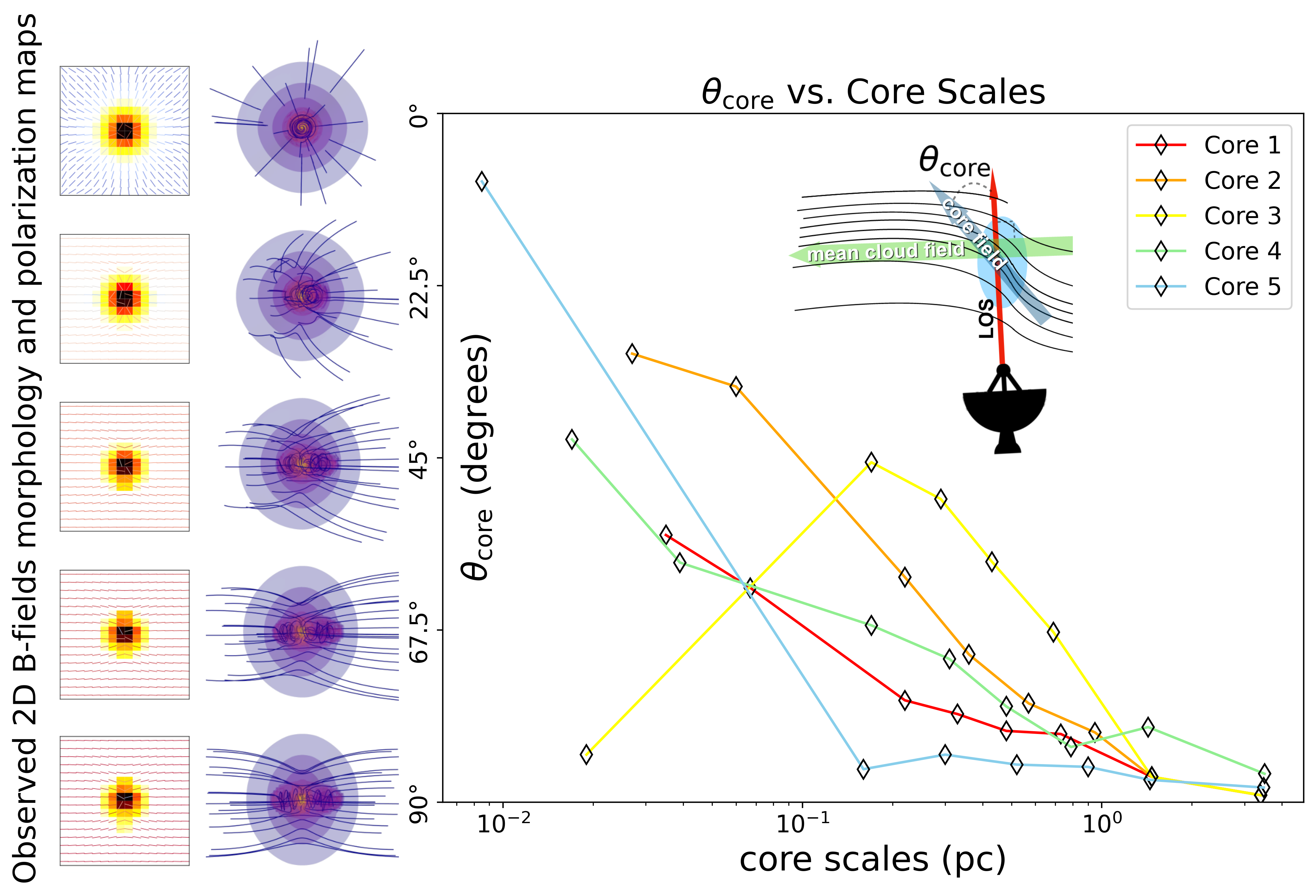}
\caption{An illustration showing the diversity of observed 2D plane-of-sky dispersion of \textit{B}-field orientations (left panel, \citealt{2024ApJ...966..237W}), made possible by the various 3D angle $\theta_{\rm core}$ on small scales for different cores in a simulated molecular cloud (right panel, \citealt{2019ApJ...871...98Z}).  On the right panel, the five cores are identified from a simulation at increasing levels of volume density thresholds and corresponded decreasing spatial scales. At a specific scale, the angle $\theta_{\rm core}$ is calculated as the angle between our LOS and the density weighted average magnetic field direction within that threshold.
Note that the left panel and right panel are from two different simulations, so the 2D magnetic fields on the left do not strictly one-to-one correspond to the cores on the right, and are only for qualitative illustrational purpose.}
\label{fig:wang2024}
\end{figure*}

The right part of Figure \ref{fig:wang2024} \citep{2024ApJ...966..237W} further illustrates that the mean magnetic fields of different cores can have a wide range of (3D) alignments ($0^\circ$ to  $90^\circ$) relative to the mean field of their parental cloud \citep{2019ApJ...871...98Z}.  This misalignment angle arises because core turbulence can be slightly super-Alfv\'{e}nic even within a globally sub-Alfv\'{e}nic cloud \citep{Cao_2023, Yuan_2025}, as we mentioned in the Introduction. This inherent variation in core-field alignment is not only the origin of the effects we described in Figure \ref{fig:dense_cores} but also implies that the angle between the LOS and an individual core's mean field varies from core to core. 

It should also be noted from Figure \ref{fig:wang2024} that while the five dense cores on the right panel have their \textit{B}-fields aligned on the large scale, the \textit{B}-field orientations gradually diverge for each core as we probe into smaller and smaller scales, again because sub-Alfv\'{e}nic turbulence turns into super-Alfv\'{e}nic turbulence on the scale of dense cores. Since higher volume densities are naturally associated with smaller scales as the result of compression and contraction, we thus see that the core-cloud field alignment is also scale- and density-dependent.

\subsection{Field Dispersion (\enquote{\texorpdfstring{$\sigma$}{Lg}} in Table \ref{tab:parameter_definitions}) is Perspective-and Density-Dependent }

We illustrate this concept using results from previous work. Due to the 3D misalignment discussed in Section \ref{sec:2.1}, the observed polarization dispersion varies significantly among cores. The left panels of Figure \ref{fig:wang2024} \citep{2024ApJ...966..237W} present several viewing angles and their corresponding polarimetry maps for a simulated cloud core. It is evident that a viewing angle closer to the mean magnetic field direction yields a higher measured dispersion in the polarimetry map. This occurs because the ordered component of the field (i.e., the component aligned with the mean field) contributes little to the polarization signal when observed along the mean-field direction. \textit{We emphasize that even with a fixed line of sight (LOS) toward a cloud, the LOS-field offset can differ from core to core, as the 3D core-cloud field offsets vary across cores (Figure \ref{fig:wang2024}; right).}

The dependence of field dispersion on density has received less attention, but it can be inferred from the inverse correlation between velocity dispersion and polarization fraction \citep{2009ApJ...704..891L, thesis}. This relationship arises because unresolved field dispersion tends to manifest as a lower polarization fraction, while velocity dispersion typically increases with density. We were the first to predict the correlation between velocity dispersion and density \citep{2019ApJ...871...98Z, 2021Galax...9...41L, Yuan_2025}, a trend that has also been confirmed by recent observations (\citealt{2023ASPC..534..193P}; Fig. 2).

\section{Simulations} \label{sec:simulation}

We follow the methods presented in \cite{Cao_2023} for creating sub-Alfv\'{e}nic, ideal-MHD simulations of molecular clouds that are in line with observational constraints \citep{2009ApJ...704..891L, 2015Natur.520..518L}. The simulation occurs in a periodic cubic domain with a size $L_{\rm box}$  of 4.8 parsec, uniformly divided into $960^3$ voxels so that each voxel has a side length of 0.005 parsec. This domain contains a molecular cloud that initially has a constant density of $n_{\mathrm{ini}} = 300\ {\rm H_2}\ \mathrm{cm}^{-3}$ and is maintained at a steady temperature of $T=10 \mathrm{~K}$ under isothermal conditions, with a mean molecular weight of 2.36. A uniform magnetic field of $14.4\ \mu \mathrm{G}$ is applied throughout the cloud in the $z$-direction. Consequently, the sound speed is $c_s=187.7 \mathrm{~m} \mathrm{~s}^{-1}$, the ratio of thermal to magnetic pressure $\beta$ is 0.05, and the mass-to-flux ratio of the whole domain is 2:
\begin{equation} \label{mass-to-flux}
   \displaystyle \frac{M}{M_{\Phi}} =  \displaystyle M \cdot \frac{2 \pi \sqrt{G}}{\Phi}  =  2.  
\end{equation}
Here $M$ is the mass of the cloud, $M_{\Phi}$ is the magnetic critical mass, and ${\Phi}$ is the magnetic
flux. For a more detailed description of the simulation setup, we refer readers to Section 2 and Appendix in \cite{Cao_2023}.

Five MHD simulations are carried out using ZEUS-MP \citep{2006ApJS..165..188H}. Each simulation is divided into two phases: Phase I focuses on driving turbulence with no self-gravity, and Phase II allows the turbulence to decay, with self-gravity turned on so that dense cores may form. The simulation snapshots that we employed in the subsequent analysis are all from the late stage snapshots in Phase II, on average at around 1.32 Myr after the end of Phase I in each of these simulations  (in \citealt{Cao_2023} Phase II lasts 2 Myr). During Phase I, turbulence is induced in a divergence-free (solely solenoidal) manner, with kinetic energy of $c_s^2 / 2$ per unit mass being injected every 0.01 million years 
following the turbulence-driving method described by \cite{2017ApJ...836...95O}. To make sure that we reach turbulence energy saturation at the end of the driving stage in Phase I, Phase I lasts relatively long time, ranging from 7.45 Myr to 11.26 Myr for the five simulations.  At the end of Phase II, the average (across the five simulations) sonic Mach number of turbulence is $\mathcal{M}_s =  \sigma_v / c_s \sim 3.82$, and the average eddy turnover time on the cloud scale is $t = L_{\rm box} /  \sigma_v \sim 6.56$ Myr.
\footnote{The numbers reported in this section involving the velocity dispersion $\sigma_v$ and the mean \textit{B}-field magnitude are all density-weighted.}
Figure \ref{fig:dense_cores} provides a visualization of selected dense cores and their surrounding magnetic fields from simulation seed 7028, taken from a late stage snapshot towards the end of Phase II in that simulation.


While the five simulations employed different random seeds so that each driving step is different in different simulations, we ensure that the five simulations exhibit the same turbulence spectrum slope upon reaching turbulence energy saturation at the end of Phase I, and that they are all sub-Alfv\'{e}nic (with $\langle \mathcal{M}_A \rangle_{\rm cloud} \sim 0.56$ at the end of Phase II across our five simulations; this number is calculated using the velocity dispersion $\sigma_v$ of the entire cloud, as well as the mean \textit{B-}field magnitude $\sqrt{\langle B^2\rangle}$ averaged over the whole cloud). The purpose of using these different simulations is to check the robustness of our method: as long as we use an overall sub-Alfv\'{e}nic simulation, any other features specific to that particular simulation (such as its velocity or density field) should not affect our main conclusion.

\section{Methods} \label{sec:methods}

\subsection{List of notations}

See Table \ref{tab:parameter_definitions} for a list of notations used in this paper and their explanations.

\begin{table}[ht]
    \centering
    \caption{Notation Definitions}
    \begin{tabular}{@{}ll@{}}
        \toprule
        \textbf{Notation} & \textbf{Description} \\ \midrule
        
        $\Delta \theta_B$ & 
        \begin{minipage}[t]{0.75\textwidth}
        Magnetic field offset. In \cite{Wu_2024}, it is the angle between a polarization half-vector probed by JCMT and that probed by Planck \citep{2015}; in our projected simulations it is defined for each pixel as the angle between the direction of the projected \textit{B}-field relative to the $z$-axis in the simulation (the direction of the initial \textit{B}-field).
        \end{minipage} \\[9ex]

        $\delta \theta_B$ & 
        \begin{minipage}[t]{0.75\textwidth}
        In \cite{Wu_2024} it is defined similar to $\Delta \theta_B$, but on a smaller scale as the angle between a polarization half-vector probed by CARMA and that probed by JCMT. In this paper, to match the way we define offsets in simulations, we define $\delta \theta_B$ as the angle between a polarization half-vector probed by CARMA and that probed by Planck, so that the observed offsets on CARMA and JCMT scales are all defined relative to the largest Planck scale.
        \end{minipage} \\[13ex]

        $\mu$ & 
        \begin{minipage}[t]{0.75\textwidth}
            The circular mean of a distribution of \textit{B}-field offsets \citep{MARDIA197218}. Note that our offsets are distributed between $-90^\circ$ and $90^\circ$ (see Section \ref{sec:4.2}).
        \end{minipage} \\[4ex]

        $\sigma$ & 
        \begin{minipage}[t]{0.75\textwidth}
        The circular standard deviation of a distribution of \textit{B}-field offsets \citep{MARDIA197218}.
        \end{minipage} \\[2ex]

        $\theta$ &\begin{minipage}[t]{0.75\textwidth}
        The polar angle in the spherical coordinate system. Since in our simulations the initial \textit{B}-field direction is along the $z$-axis, when we use projections the angle $\theta$ describes the angle between our line-of-sight (LOS) and the initial \textit{B}-field direction.  See Figure \ref{fig:coodinates} for an illustration.
        \end{minipage} \\[7ex]
        
        $\phi$ & The azimuthal angle in the spherical coordinate system. \\
        
        \bottomrule
    \end{tabular}
    \label{tab:parameter_definitions}
\end{table}

\begin{figure*}[ht]
\centering
\includegraphics[scale = 1.65]{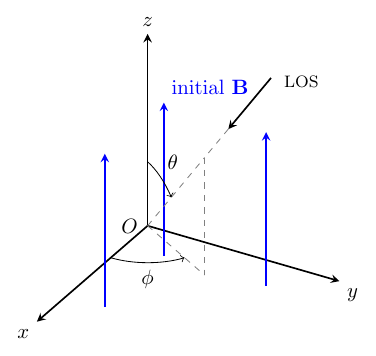}
\caption{The spherical coordinate system used in this work.}
\label{fig:coodinates}
\end{figure*}

\subsection{Characterizing polarization measurements}\label{sec:4.2}

In Figure 2 from \cite{Wu_2024}, the distribution of \textit{B}-field offsets $\delta \theta_B$ and $\Delta \theta_B$ are used to describe whether the \textit{B}-fields are well-aligned (small mean $\mu$, small standard deviation $\sigma$) or more randomly aligned (large $\mu$, large $\sigma$). We wish to use compact summary statistics that could capture some key features in the morphology of \textit{B}-fields, and we choose to use the mean $\mu$ and standard deviation $\sigma$ of the distribution of 2D \textit{B}-field offsets as our metric. We could have used more variables such as velocity dispersion or estimated Alfv\'{e}n Mach number to characterize the observations, but since \cite{Wu_2024} and many other observers make conclusions about the relative significance of turbulence and \textit{B}-fields in observed molecular clouds based on the 2D offsets distributions, we choose to stick to the $\mu$ and $\sigma$ for this paper. The aim would be to assess whether the ($\mu$, $\sigma$) measured from the three cores in the Orion ISF are sufficient to provide evidence against specific cloud models. In particular, we examine whether the observational results of \cite{Wu_2024} can provide evidence against the sub-Alfvénic cloud model adopted in \cite{Cao_2023} and \cite{Yuan_2025}.

\begin{figure*}[ht!]
\centering
\includegraphics[scale = 0.4]{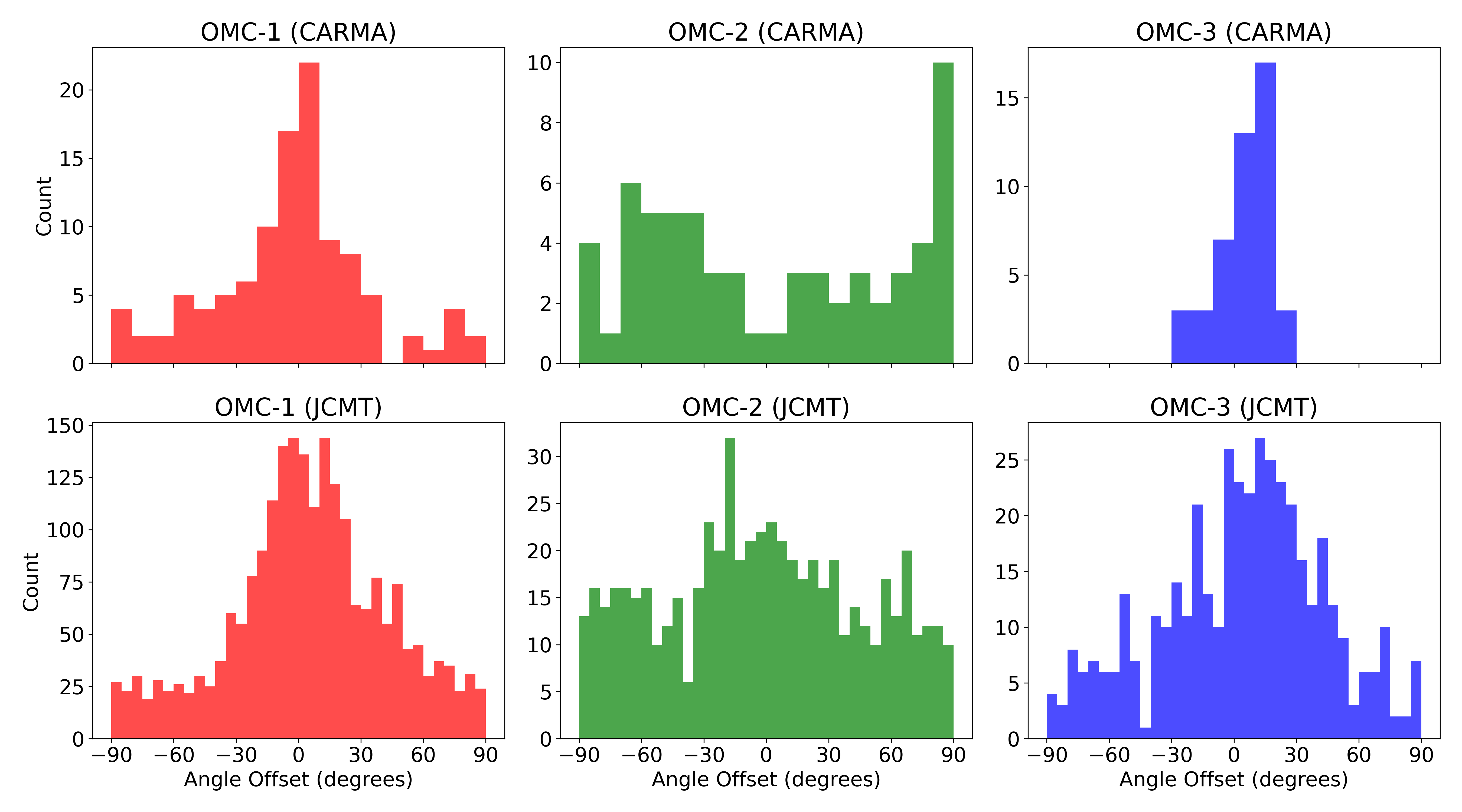}
\caption{The six observed sets of offset angles from Figure 2 in \cite{Wu_2024}, but are now distributed between $-90^\circ$ and $90^\circ$ instead of between $0^\circ$ and $90^\circ$ as in the original Figure. Note that these distributions are periodic with $180^\circ$.}
\label{fig:offsets}
\end{figure*}

Thus, for the 6 \textit{B}-field offsets $\delta \theta_B$ and $\Delta \theta_B$ distributions from CARMA, JCMT, and Planck polarization maps in Figure \ref{fig:offsets}, we calculate their mean $\mu$ and standard deviation $\sigma$, and obtained 6 sets of data-points in the ($\mu$, $\sigma$) space (Figure \ref{fig:data_points}) to summarize the polarization measurements of OMC-1, OMC-2 and OMC-3. Notice that to match the way we define offsets in simulations, here we define $\delta \theta_B$ a little differently (as described in Table \ref{tab:parameter_definitions}) compared to the definition in \cite{Wu_2024}, though this does not significantly change the resulted offsets.

\subsection{Processing simulation data} \label{subsec:projections}

Similar to how we process the observational data, we wish to use ($\mu$, $\sigma$) as a summary statistics to describe our MHD simulations as well. But before we can do that, we need to project our 3D MHD simulations to 2D maps of \textit{B}-field so that we may calculate our summary statistics. To achieve this, we use the \texttt{off\_axis\_projection()} function from the \texttt{yt} \citep{2011ApJS..192....9T} package. In order to obtain 2D \textit{B}-field maps, we first choose a specific LOS direction $\hat{\mathbf{e}}$ for viewing our 3D simulation, which is equivalent to choosing a specific pair of polar angle $\theta$ and azimuthal angle $\phi$ in the spherical coordinate system. Then we decompose the simulated \textit{B}-fields into Stokes parameters ($q$, $u$) (following \citealt{2024ApJ...966..237W}), weighting them by volume densities, and finally apply the \texttt{yt} function to project them to a 2D polarization map.

 It should be noted here that weighting by density is equivalent to assuming a constant grain alignment efficiency. While there are theories suggesting that the efficiency of dust grain alignment should decrease with density (e.g., \citealt{2015ARA&A..53..501A}), largely motivated by the polarization hole (PH) phenomenon, other studies show that PH is independent of grain alignment efficiency (e.g., \citealt{2024ApJ...966..237W}). For simplicity, we follow the latter to emphasize on what projection alone can achieve in the diversity of polarization morphology. 
Another caveat to note is that the way we use the \texttt{yt} function \texttt{off\_axis\_projection()} for projection is equivalent to assuming that the medium is optically thin for dust emission. While it may not be as accurate as a full radiative transfer treatment (e.g., \citealt{2016A&A...593A..87R}), we consider the method to be a good approximation in the (sub-)millimeter regime where dust emission is typically optically thin (e.g., \citealt{1983QJRAS..24..267H}).

\begin{figure*}[ht!]
\centering
\includegraphics[scale = 0.5]{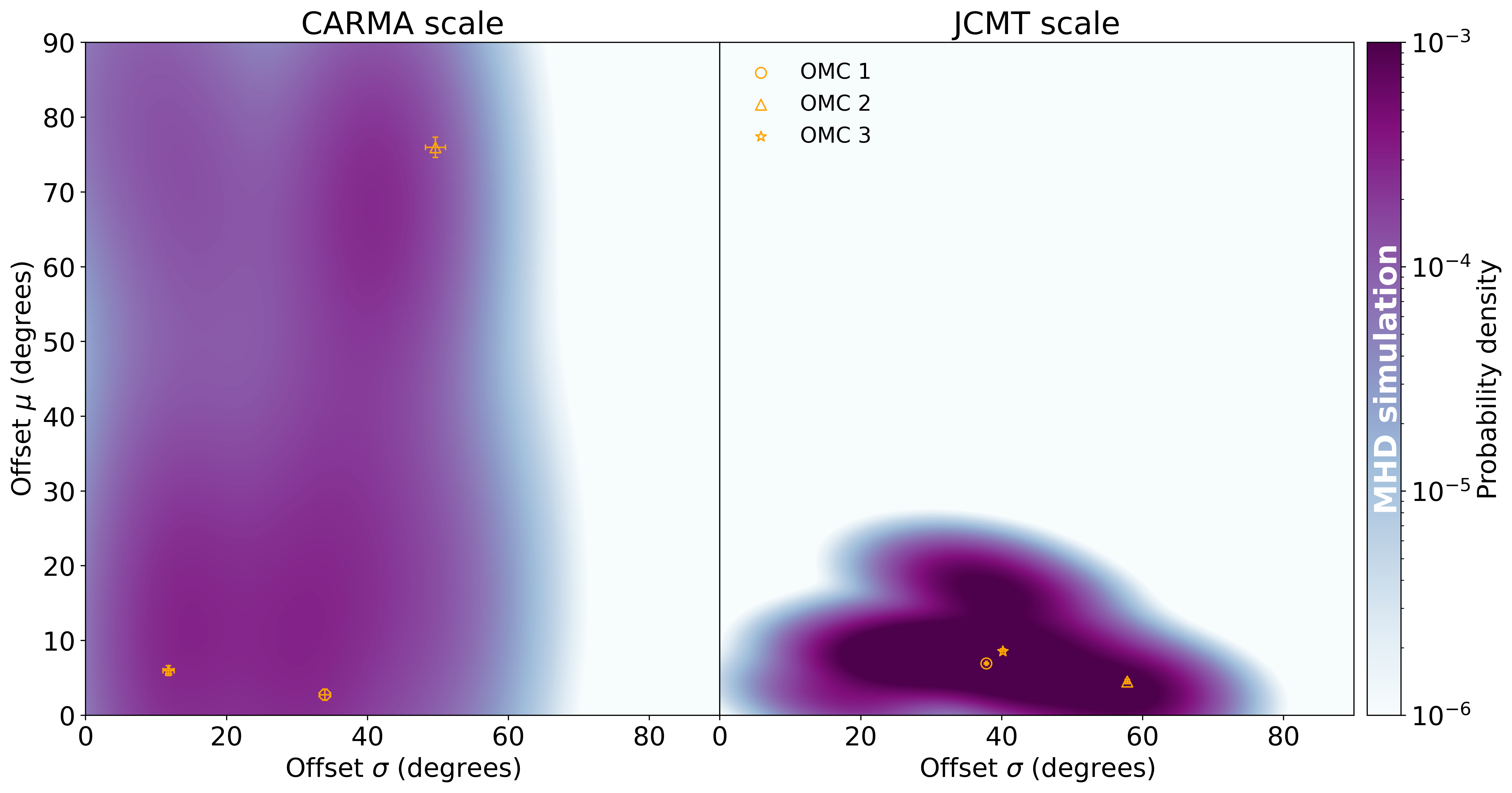}
\caption{The six observed data-points of Orion ISF \citep{Wu_2024} in ($\mu$, $\sigma$) space and the best fit probability distributions (at two different scale ranges) generated from a MHD simulation at $\theta = 35^\circ$, $\phi = -18^\circ$. The lower limit of the colorbar is set at $10^{-6}$ to allow more contrast between regions with high and low probability densities. The errorbars of the CARMA and JCMT data-points are calculated by error propagation from the polarization angles uncertainties in the CARMA data, and from the Stokes $Q, U$ uncertainties in the JCMT data. We also take absolute value for offset $\mu$; e.g., $\mu = -60^\circ$ is the same as $\mu = 60^\circ$.}
\label{fig:data_points}
\end{figure*}

\begin{algorithm}[ht!]
\SetAlgoLined
\DontPrintSemicolon
\caption{Derive a probability distribution in ($\mu$, $\sigma$) space by projecting MHD simulations}
\label{alg:contour_analysis}
\Begin{
    Define projection angles $\theta$ and $\phi$ \;
    Define a series of contour levels based on volume densities
    
    \For {core $c \in \{1, 2, \ldots, 10\}$}{
        \For{contour $k \in \{1, 2, \ldots, \text{max\_contour\_number}\}$}{
            Project 3D core $c$ within contour $k$ to 2D map using \texttt{off\_axis\_projection()} in \texttt{yt} \;
            
            \For{pixel $(i,j)$ in 2D map}{
                Calculate the \textit{B}-field offset $\Delta\theta_B^{(i,j)}$ relative to $z$-axis (initial \textit{B-}field direction) \;
            }
            Compute $\mu_c^{(k)} \leftarrow \text{mean}(\Delta\theta_B)$ \;
            Compute $\sigma_c^{(k)} \leftarrow \text{std}(\Delta\theta_B)$ \;
        }
    }
    
    Aggregate all pairs $(\mu_c^{(k)}, \sigma_c^{(k)})$ from different cores and contours \;
    Convert the discrete $(\mu_c^{(k)}, \sigma_c^{(k)})$ distribution to a continuous probability distribution $f(x)$ in $(\mu, \sigma)$ space using Kernel Density Estimation (KDE) \;
}
\end{algorithm}

In our MHD simulation of molecular clouds, we define a \enquote{core} as a region in the simulation cube where all the voxels in the region are above a certain volume density threshold. We take this threshold to be 20 times the initial volume density, i.e. $n_{\rm threshold} = 20\ n_{\rm ini} = 6 \times 10^3\ {\rm H_2}\ \mathrm{cm}^{-3}$. In a given simulation, we first find the 10 largest cores (in terms of number of pixels contained within the threshold contour) based on this threshold; these cores are then used to derive a probability distribution in the ($\mu$, $\sigma$) space, following the procedure outlined in Algorithm \ref{alg:contour_analysis}. Figure \ref{fig:data_points} shows the six observed data-points from \cite{Wu_2024} in the ($\mu$, $\sigma$) space, along with an example of probability distribution obtained from a simulation at $\theta = 40^\circ$ in the background. In Algorithm \ref{alg:contour_analysis}, we chose to define 30 contour levels based on volume densities, which are 30 numbers uniformly spaced in logarithm scale from $6 \times 10^3\ {\rm H_2}\ \mathrm{cm}^{-3}$ to $1 \times 10^5\ {\rm H_2}\ \mathrm{cm}^{-3}$. This range of volume densities roughly corresponds to a 2D projected scale range of 0.012 pc to 0.9 pc for our simulated cores.


For the Kernel Density Estimation (KDE) in Algorithm \ref{alg:contour_analysis}, we applied the \texttt{scipy} module \texttt{scipy.stats.gaussian\_kde}, which uses Gaussian kernels and determines the estimator bandwidth automatically using Scott's rule by default \citep{2020SciPy-NMeth}. Now for the observed regions probed by JCMT and CARMA in Orion ISF, we estimate their scale range to be from 0.06 pc to 0.2 pc for CARMA, and from 0.71 pc to 1.06 pc for JCMT. We then apply the two scale range filters to the projected contours, to construct two separate KDE distributions for the JCMT and CARMA datasets (as shown in Figure \ref{fig:data_points}), and combine the results from two datasets later (see Section \ref{subsec:stats}).

\subsection{Circular statistics}

Also shown in Figure \ref{fig:data_points} are the ($\mu$, $\sigma$) values derived from Orion ISF as in \cite{Wu_2024}. If a particular projection of the simulation model yields a probability density distribution that is statistically consistent with the observations (Section \ref{subsec:stats}), then the observations cannot provide enough evidence against that model. To amplify potential differences between the models and the observations, we adopt \textit{Circular Statistics}. The mean angle $\mu$ obtained from circular statistics is identical to that derived from the conventional Stokes-mean polarization. In contrast, whereas the conventional dispersion $\sigma$ saturates at approximately 52 degrees, the dispersion derived from circular statistics (which is described in detail below) has no upper limit. This provides a larger dynamic range and thus greater sensitivity to subtle differences.
In circular statistics \citep{MARDIA197218}, the mean is defined as
$$\mu = \frac{1}{2} \text{arctan2}( \bar{U}, \bar{Q}),$$
where $\bar{Q}$ and $\bar{U}$ are the averaged Stokes parameters. And the standard deviation is defined as
$$ \sigma = \frac{1}{2} \sqrt{-2 \ln R}, $$ where $ R $ is the length of the mean resultant vector with $R = \sqrt{ \bar{Q}^2 + \bar{U}^2}$. Here perfect alignment of the angle distributions would have $R=1$ and random scattering (uniform distribution) leads to $R=0$.

\subsection{Statistical methods for comparing observations with simulations\label{subsec:stats}}

To quantify the similarity between observations and simulations, we use two complementary statistical approaches. First, maximum likelihood estimation identifies the projection angle $(\theta,\phi)$ for which the observed $(\mu,\sigma)$ values are most likely under the simulation-derived distributions. Second, hypothesis testing asks whether the observed $(\mu,\sigma)$ values provide sufficient evidence to reject the simulation model at a given projection angle.

\subsubsection{Maximum likelihood estimation}

For each projection direction, specified by the polar angle $\theta$ and azimuthal angle $\phi$, we construct two scale-dependent probability density functions in the $(\mu,\sigma)$ plane: $f_{\rm JCMT}(\mu,\sigma\mid\theta,\phi)$ for the JCMT-scale data and $f_{\rm CARMA}(\mu,\sigma\mid\theta,\phi)$ for the CARMA-scale data. These distributions are obtained from separate KDEs applied to the projected simulation outputs after imposing the corresponding observational scale ranges, as described in earlier sections.

Let the observed JCMT-scale measurements be written as $(\mu_{{\rm JCMT},i},\sigma_{{\rm JCMT},i})$ for $i=1,\ldots,N_{\rm JCMT}$, and the observed CARMA-scale measurements as $(\mu_{{\rm CARMA},j},\sigma_{{\rm CARMA},j})$ for $j=1,\ldots,N_{\rm CARMA}$. For the Orion ISF data used here, $N_{\rm JCMT}=3$ and $N_{\rm CARMA}=3$. Assuming that the points within each observational scale can be treated as independent samples from the corresponding scale-dependent distribution, and that the two scale-specific datasets are conditionally independent for a fixed projection geometry, the joint likelihood is
\begin{equation} \label{likelihood}
\begin{split}
L(\theta,\phi) =
&\left[\prod_{i=1}^{N_{\rm JCMT}}
 f_{\rm JCMT}(\mu_{{\rm JCMT},i},\sigma_{{\rm JCMT},i}\mid\theta,\phi)\right] \\
\times &
\left[\prod_{j=1}^{N_{\rm CARMA}}
 f_{\rm CARMA}(\mu_{{\rm CARMA},j},\sigma_{{\rm CARMA},j}\mid\theta,\phi)\right].
\end{split}
\end{equation}

Different projection angles generally give different likelihoods for the same observed data. We therefore estimate the most likely viewing geometry by finding the pair $(\theta,\phi)$ that maximizes $L(\theta,\phi)$ in Equation \ref{likelihood}. Equivalently, we minimize the negative log-likelihood,
\begin{equation}
\begin{split}
-\ln L(\theta,\phi) =
&-\sum_{i=1}^{N_{\rm JCMT}}
\ln f_{\rm JCMT}(\mu_{{\rm JCMT},i},\sigma_{{\rm JCMT},i}\mid\theta,\phi) \\
&-\sum_{j=1}^{N_{\rm CARMA}}
\ln f_{\rm CARMA}(\mu_{{\rm CARMA},j},\sigma_{{\rm CARMA},j}\mid\theta,\phi).
\end{split}
\end{equation}

If the prior distribution of projection angles is taken to be uniform over the explored angular grid, maximizing the likelihood is equivalent to maximizing the posterior probability of the projection angle. Thus the best-fit viewing geometry is
\begin{equation}
(\theta,\phi)_{\rm best}
= \arg\max_{\theta,\phi} L(\theta,\phi)
= \arg\min_{\theta,\phi}[-\ln L(\theta,\phi)].
\end{equation}

In practice, we first evaluate the negative log-likelihood on a coarse set of approximately evenly spaced viewing directions generated with a Fibonacci lattice (e.g., \citealt{Gonz_lez_2009}). For each trial direction, the simulation is projected, the JCMT- and CARMA-scale KDEs are constructed separately, and the likelihood of the observed $(\mu,\sigma)$ values is evaluated. The grid point with the smallest negative log-likelihood is then used as the starting point for a local numerical minimization using the Nelder--Mead method implemented in \texttt{scipy.optimize.minimize} \citep{2020SciPy-NMeth}. The angles sampled during the coarse search and subsequent local optimization form the angular grid shown in the results section (see Figure \ref{fig:nll} for a visualization).

\subsubsection{Hypothesis test}

The maximum-likelihood analysis identifies which projection angles best match the observations, but the likelihood value alone does not determine whether the observed data are statistically compatible with the simulation-derived distributions. We therefore perform hypothesis testing to assess whether the observed $(\mu,\sigma)$ values provide sufficient evidence against the adopted sub-Alfv\'enic simulation model at each projection angle.

For a specified viewing geometry $(\theta,\phi)$, the null hypothesis is that the observed JCMT-scale and CARMA-scale points in the $(\mu,\sigma)$ plane are drawn from their corresponding simulation-derived distributions:

\begin{equation}
H_0(\theta,\phi):\quad
(\mu,\sigma)_{{\rm JCMT},{\rm obs}} \sim f_{\rm JCMT}(\mu,\sigma\mid\theta,\phi)
\quad \text{and} \quad
(\mu,\sigma)_{{\rm CARMA},{\rm obs}} \sim f_{\rm CARMA}(\mu,\sigma\mid\theta,\phi).
\end{equation}

The alternative hypothesis is that at least one of the two observed scale-dependent datasets is inconsistent with its corresponding distribution:

\begin{equation}
H_1(\theta,\phi):\quad
(\mu,\sigma)_{{\rm JCMT},{\rm obs}} \not\sim f_{\rm JCMT}(\mu,\sigma\mid\theta,\phi)
\quad \text{or} \quad
(\mu,\sigma)_{{\rm CARMA},{\rm obs}} \not\sim f_{\rm CARMA}(\mu,\sigma\mid\theta,\phi).
\end{equation}

At each projection angle, we test the JCMT-scale and CARMA-scale datasets separately. For each scale, we compare the observed points in the $(\mu,\sigma)$ plane with samples drawn from the corresponding KDE distribution. We use two non-parametric two-sample procedures: the energy-distance permutation test and the maximum mean discrepancy (MMD) permutation test. Their mathematical definitions and permutation procedures are given in \hyperref[sec:appendix]{Appendix A}.

Each procedure returns one $p$-value for the JCMT-scale comparison and one $p$-value for the CARMA-scale comparison. We denote these by $p_{\rm JCMT}$ and $p_{\rm CARMA}$, respectively. To summarize the joint evidence across the two observational scales, we combine the two $p$-values using Fisher's method \citep{fisher1970statistical},

\begin{equation}
\chi^2_F = -2\left[\ln(p_{\rm JCMT}) + \ln(p_{\rm CARMA})\right].
\end{equation}

If the two $p$-values were independent under $H_0(\theta,\phi)$, $\chi^2_F$ would follow a chi-squared distribution with four degrees of freedom, from which the combined $p$-value can be computed. In the present application, the independence assumption is not exact because the JCMT- and CARMA-scale observations are probing the same physical regions inside a molecular cloud. Treating the two tests as independent therefore makes Fisher's method rejection-favorable: it gives the observations the best chance to reject the null hypothesis $H_0$. Consequently, any projection angle $(\theta,\phi)$ that is not rejected by this procedure would also not be rejected once the dependence between scales is treated more conservatively.

For each projection angle $(\theta,\phi)$, we reject the sub-Alfv\'enic model at that viewing geometry if the combined $p$-value is smaller than 0.05. Otherwise, we fail to reject the null hypothesis $H_0$ and regard the observed $(\mu,\sigma)$ values as statistically consistent with the projected sub-Alfv\'enic simulation at that viewing geometry. The resulting rejection and non-rejection regions are presented in Section \ref{sec:results}.

\section{Results} \label{sec:results}

\begin{figure*}[ht!]
\centering
\includegraphics[scale = 0.60]{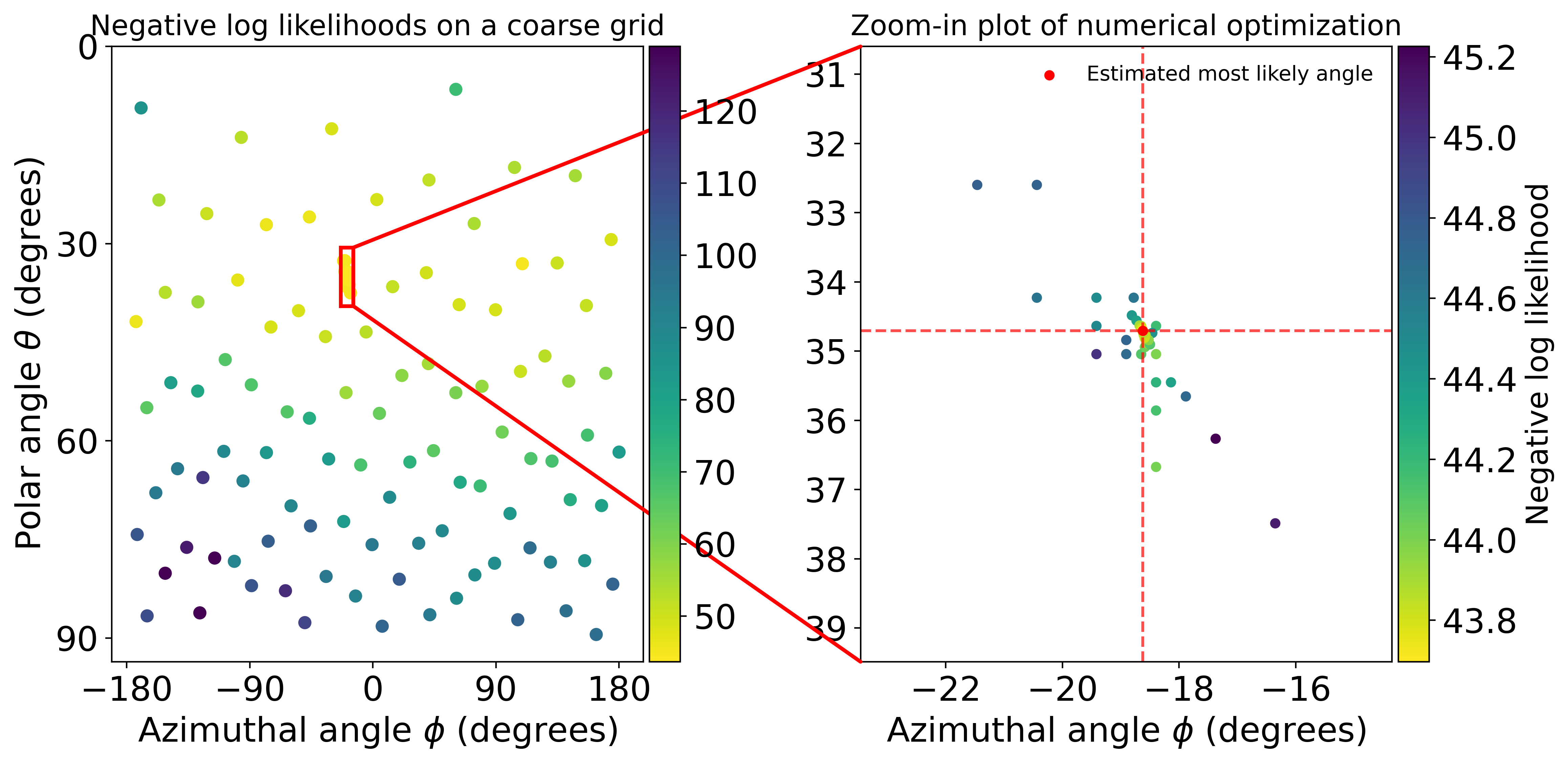}
\caption{Likelihoods for the different projections angles of the observed polarization maps over the whole ($\theta$, $\phi$) space (left panel), and in a small zoom-in region where we execute our numerical optimizations (right panel). Lower NLL value (larger likelihood) means that the observed polarization maps from Orion ISF are more likely to have those projection angles. To enhance visual contrast, in the left panel all the NLL values greater than 45 are displayed using the same color, thereby allowing finer differentiation in colors among values below 45.
\label{fig:nll}}
\end{figure*}

Figure \ref{fig:nll} shows the likelihood of the observed polarization data from the Orion ISF as a function of projection angles $(\theta, \phi)$, evaluated using probability distributions derived from a representative sub-Alfvénic MHD simulation (seed 7028). Lower negative log likelihood (NLL) corresponds to a better statistical match between the observations and the projected simulation. A clear preference emerges for projection angles around $\theta \sim 35^\circ$.

\begin{figure*}[ht!]
\centering
\includegraphics[scale = 0.6]{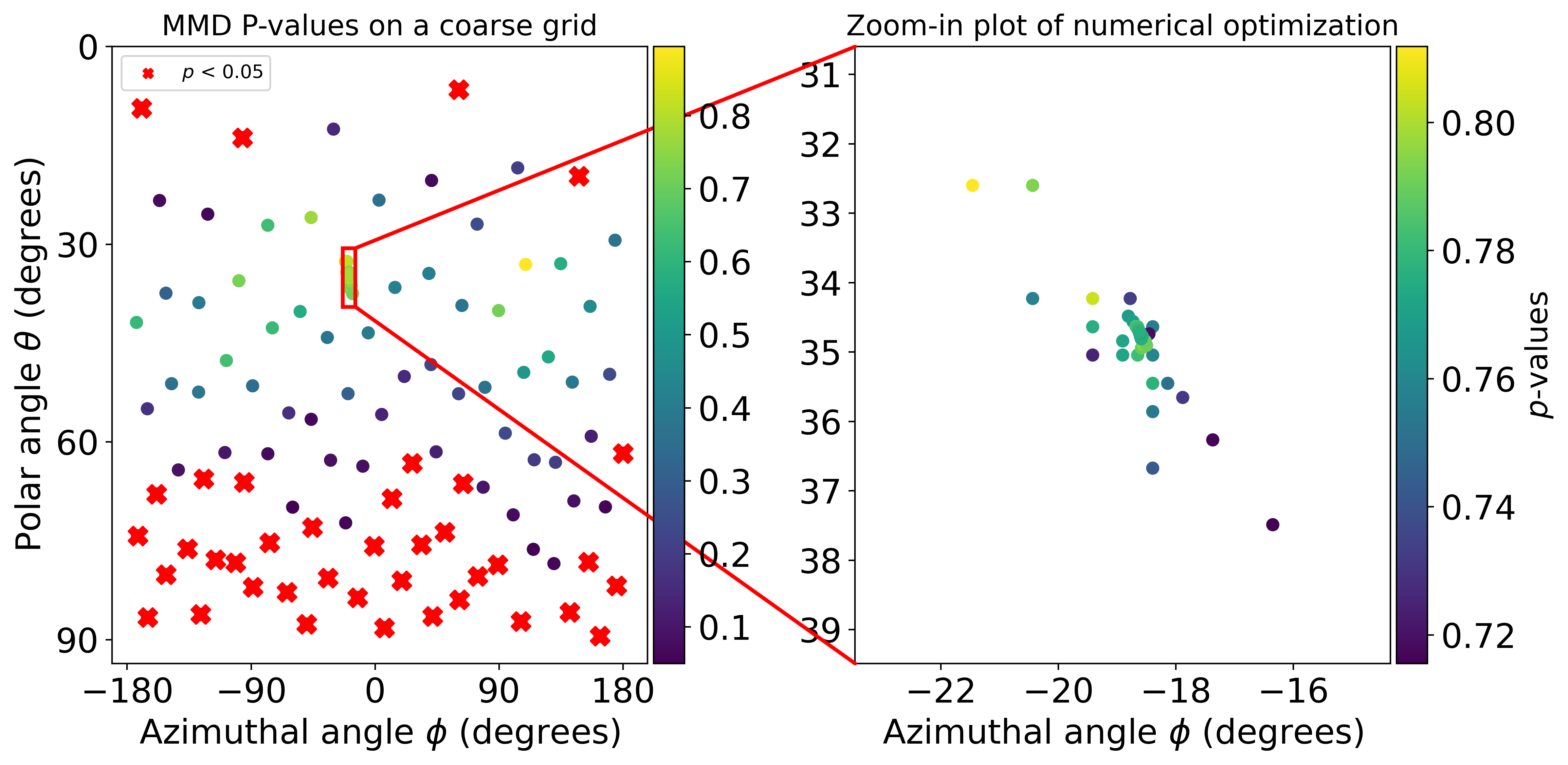}
\caption{Similar to Figure \ref{fig:nll}, but illustrating different MMD $p$-values for different projection angles instead. Angles with $p$-values smaller than 0.05 are masked with red crosses, which means that the observed polarization maps provide evidence against those projection angles. In contrast, points without red crosses indicate that our current observational dataset fail to reject those projection angles.
\label{fig:mmd-p-values}}
\end{figure*}

\begin{figure*}[ht!]
\centering
\includegraphics[scale = 0.6]{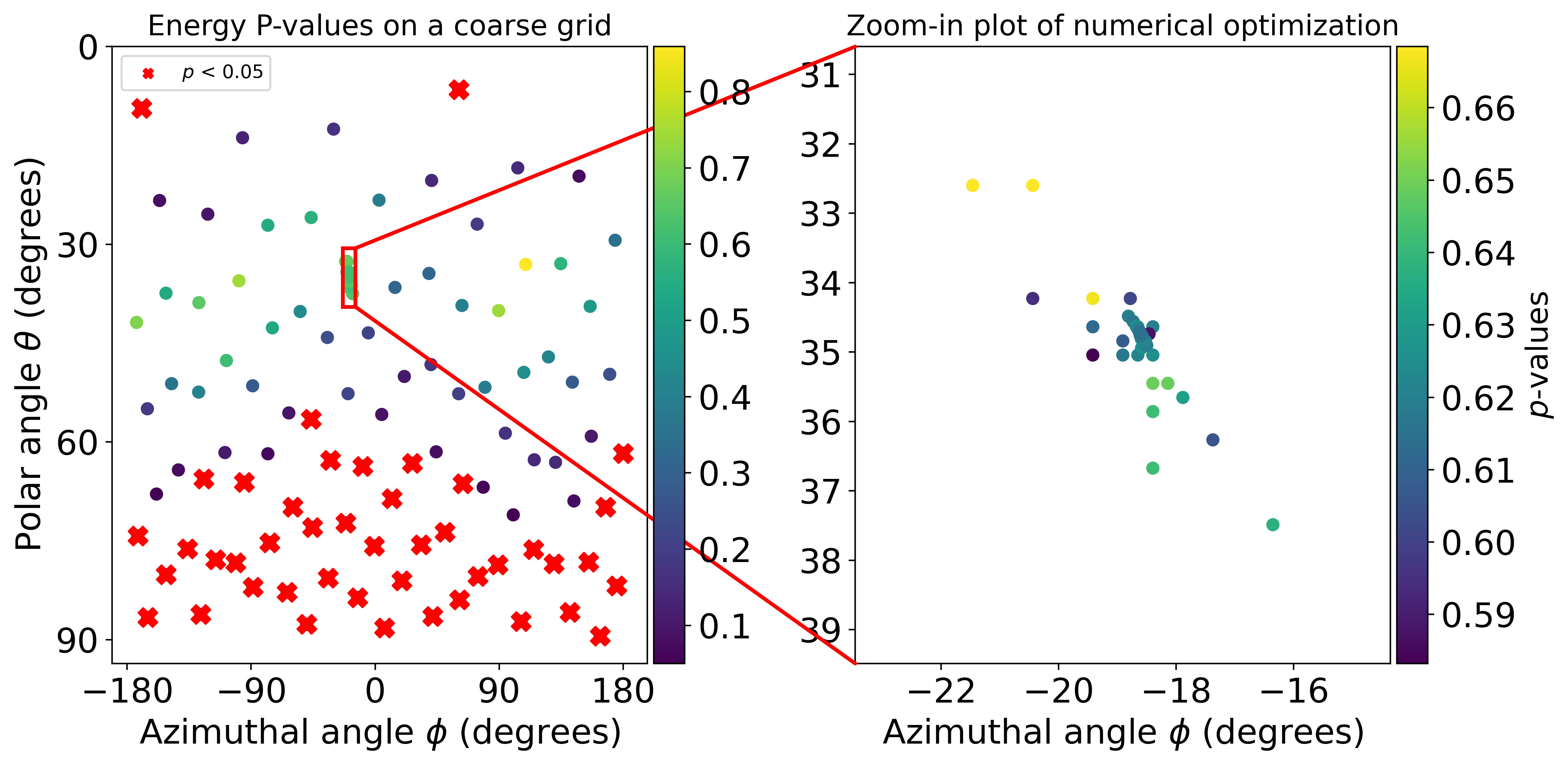}
\caption{Similar to Figure \ref{fig:mmd-p-values}, but plotting the $p$-values from the energy distance test instead.
\label{fig:energy-p-values}}
\end{figure*}

Figures \ref{fig:mmd-p-values} and \ref{fig:energy-p-values} present complementary results using hypothesis testing based on the MMD and energy distance permutation tests, respectively. Projection angles for which joint $p-$value falls below 0.05 are those for which the observed polarization statistics provide evidence against the null hypothesis that the data are drawn from the simulated distributions. In contrast, angles with larger $p-$values indicate that the current observations are statistically consistent with the sub-Alfvénic model.

The key result is that, while the observations can provide evidence against a subset of projection angles, the majority of viewing geometries cannot be rejected under the sub-Alfvénic setup. In other words, the observed $(\mu, \sigma)$ values from OMC-1, OMC-2, and OMC-3 remain statistically consistent with projected sub-Alfvénic simulations across a broad region of parameter space.

\begin{figure*}[ht!]
\centering
\includegraphics[scale = 0.6]{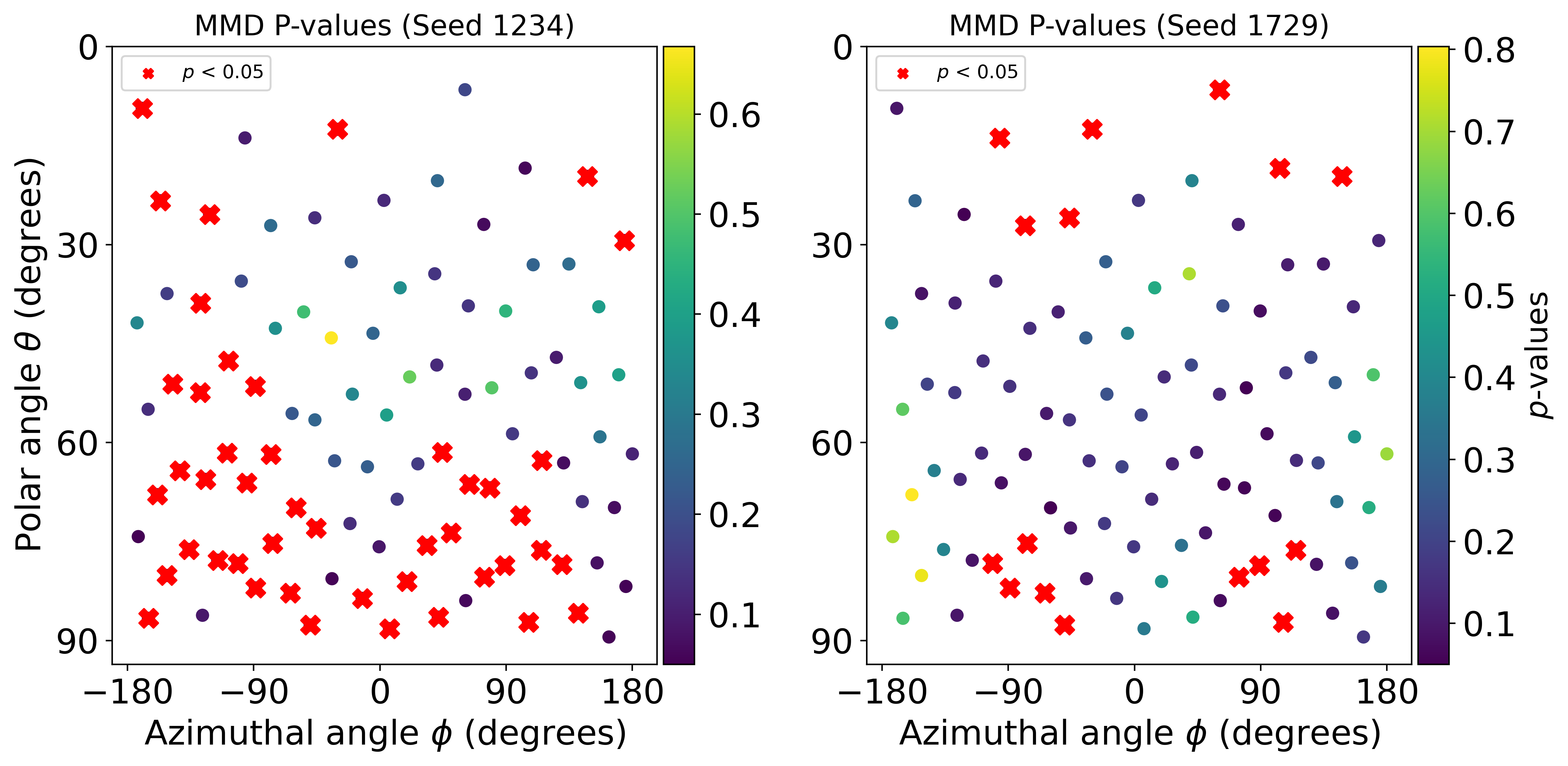}
\caption{Additional hypothesis testing results from two more simulation seeds, illustrating the MMD p-values for different projection angles.
\label{fig:more_seeds}}
\end{figure*}

To test the robustness of this conclusion, we repeated the entire analysis using two additional sub-Alfvénic MHD simulations with different turbulence realizations (Figure \ref{fig:more_seeds}, seed 1234 and seed 1729). Although the specific projection angles that are disfavored vary from simulation to simulation, all runs show the same qualitative behavior: for most projection angles, the current Orion polarization data fail to reject the sub-Alfvénic cloud model. This demonstrates that our conclusions are not sensitive to the details of individual simulations, but instead reflect a general property of how $(\mu, \sigma)$ behave under projection.
Taken together, these results show that polarization statistics based solely on $(\mu, \sigma)$ can constrain viewing geometry, but do not provide sufficient evidence against globally sub-Alfvénic cloud models given the limited number of observed cores.

\section{Discussion} \label{sec:discussion}

\subsection{Our simulation setup and results}

As described in Section \ref{sec:simulation}, our simulation uses an ideal-MHD setup, which assumes that the magnetic field lines are well coupled (\enquote{frozen-in}) to the gas motion and is a good approximation for capturing most of the behaviors of magnetic field in molecular clouds. With that said, at the scale of observed dense cores probed by JCMT and CARMA, non-ideal MHD effects such as turbulence-driven ambipolar diffusion (AD) \citep{2008ApJ...677.1151L, 2018ApJ...862...42T} could already start to play an important role. For example, \cite{2018ApJ...862...42T} estimated the ion-neutral decoupling in a starless molecular cloud located in NGC 6334 to be below 0.404 pc.  This means that AD starts to become efficient below this scale for this particular cloud, 
and thus the ideal-MHD assumption we used in this work would no longer be a good approximation. Future studies with ample computational resources (required for non-ideal MHD simulations) and a greater focus on the small scale \textit{B}-field morphology should consider including these non-ideal MHD effects carefully into their numerical setups  (e.g., \citealt{2025A&A...700A.152T, 2025RASTI...4af054C}), to more accurately reflect what we could observe from the magnetic fields in molecular clouds.

The peak volume density of cores in our simulations could be another concern. There are some estimates that put the peak volume density in Orion ISF to be above $10^6 \ {\rm H_2}\ \mathrm{cm}^{-3}$ (e.g. \citealt{2026ApJ...997..345Z}; see their Figure 9 \& Figure 12); while for the dense cores we identified in our simulations, they rarely reach such high peak volume density.
As an example, in Figure \ref{fig:dense_cores}, the two selected cores from our simulations have peak volume densities of around $3 \times 10^5 \ {\rm H_2}\ \mathrm{cm}^{-3}$ and $6.5 \times 10^4 \ {\rm H_2}\ \mathrm{cm}^{-3}$, respectively. Therefore, the peak volume density or the dynamic range of volume densities in real molecular clouds could be higher than that in our simulations.

A possible reason for why our simulations may not have reached the peak volume density estimated in real molecular clouds could be that we simply did not run the simulations for long enough. In our simulation setup, the Phase II for gravitational collapse (described in Section \ref{sec:simulation}) lasts relatively short time, and is stopped before any core in the simulation could violate the Truelove criterion that would otherwise cause artificial and unphysical fragmentation \citep{1997ApJ...489L.179T}. Although the gravitational collapsing phase did not last long enough to create dense cores with very high peak volume densities, the fact that we have already observed diverse and rich morphologies of \textit{B}-fields at our current level of peak volume density should give us confidence for our simulation setup. If we were to run a longer simulation under the same simulation setup, we should expect to see higher peak volume density and even more diverse \textit{B}-field morphologies, which should make it easier for us to find cases in the simulations that would match the real observations.

\subsection{A potential new method for probing 3D \textit{B}-field orientation?}

Although our primary goal is to assess the limitations of using $(\mu, \sigma)$ to diagnose the Alfvénic nature of molecular clouds, our results point to a different and potentially useful application. Specifically, we find that the probability distribution of $(\mu, \sigma)$ is strongly dependent on the three dimensional angle between the mean magnetic field and the line of sight, while remaining largely insensitive to the specific realization of simulation within a given sub-Alfvénic regime.

This behavior is illustrated in Appendix \ref{sec:appendix_B}, where we show $(\mu, \sigma)$ distributions aggregated over different azimuthal angles $\phi$ but fixed polar angle $\theta$. For a given $\theta$, simulations with different turbulence seeds produce remarkably similar distributions, whereas changing $\theta$ leads to systematic and substantial shifts in the $(\mu, \sigma)$ space. These results suggest that $(\mu, \sigma)$ are intrinsically sensitive to projection geometry rather than to the detailed properties of turbulence.

In this sense, polarization statistics may offer a way to probe $\theta$, provided that projection effects are modeled explicitly and statistically. However, we emphasize that such an application remains exploratory. Quantitative inference of three dimensional field orientations would require larger observational samples, additional summary statistics beyond $(\mu, \sigma)$, and extensive testing using mock observations. Nonetheless, the results presented here demonstrate that the reason that prevent $(\mu, \sigma)$ from accessing $\mathcal{M_A}$ may guide us to properly use this summary statistics.

\section{Conclusion} \label{sec:conclusion}

In this work, we interpret polarization observations of the Orion Integral-Shaped Filament (ISF) by statistically comparing commonly used polarization summary statistics with those derived from systematically projected MHD simulations. By adopting globally sub-Alfvénic simulations that naturally produce slightly super-Alfvénic dense cores, we explicitly account for both projection effects and the deviation of core scale magnetic fields from the parent cloud field.
Our results show that modest core scale field deviations (which are also consistent with previous numerical studies) are sufficient to generate a wide diversity of observed plane of sky polarization dispersions when combined with projection. Through maximum likelihood estimation and hypothesis testing, we find that the observed $(\mu, \sigma)$ values of OMC-1, OMC-2, and OMC-3 are statistically consistent with sub-Alfvénic cloud models over a broad range of viewing angles, while only a limited subset of projection geometries can be rejected. Repeating the analysis with multiple simulations further demonstrates that these conclusions are independent of the specific realization of simulations.

Importantly, our results do not constitute evidence in favor of sub-Alfvénic cloud models. Rather, they demonstrate a fundamental limitation of polarization-based diagnostics that rely solely on the summary statistics $(\mu, \sigma)$. This limitation does not imply that polarization observations lack constraining power on the Alfvén Mach number $\mathcal{M_A}$, just it is not the focus of the present work. If the explicit goal is to discriminate the Alfvén Mach number, this framework can be naturally extended by incorporating globally super-Alfvénic simulations and performing controlled statistical comparisons across models with different $\mathcal{M_A}$. Such an approach would allow future studies to assess whether super- or sub-Alfvénic clouds provide a better match to observations.

Last but not least, our analysis shows that the dependence of $(\mu, \sigma)$ on projection geometry suggests another potential application of polarization statistics: probing the three dimensional orientation of cloud scale magnetic fields relative to the line of sight.

\begin{acknowledgments}
We would like to thank Shibo Yuan for his help and insightful discussions on the numerical simulations used in this work. This research is supported by General Research Fund grants from the Research Grants Council of Hong Kong: No. 14304616, 14305717, 14306323, 14307225, and 14307222; by Collaborative Research Fund grant No. C4012-20E; and by CUHK Science Faculty Collaborative Research Impact Matching Scheme (2025).
\end{acknowledgments}

%


\software{\texttt{PyVista} \citep{2019JOSS....4.1450S}, \texttt{SciPy} \citep{2020SciPy-NMeth},
\texttt{yt} \citep{2011ApJS..192....9T}
}

\bibliography{sample7}{}
\bibliographystyle{aasjournalv7}


\appendix 

\section{Hypothesis testing} \label{sec:appendix}

Here we detail the mathematical descriptions of the methods we used in hypothesis testing. We have two sets of potentially dependent observed two-dimensional data points $X_1$ and $X_2$ of sizes $N_1$ and $N_2$, respectively. In the main text, these two datasets correspond to the JCMT-scale and CARMA-scale observations, respectively. The goal is to determine whether the distribution of each observed data set is statistically consistent with the corresponding distribution derived from our MHD simulations. We use kernel density estimators (KDEs) on the simulated data, yielding a smooth approximation of the underlying probability densities, $f_1(x)$ and $f_2(x)$. Then, this question can be formalized as a test of whether the empirical distributions $F_1(x)$ and $F_2(x)$ of the observed data match their respective distributions $f_1(x)$ and $f_2(x)$ derived from simulations. We test the joint null hypothesis

{
\[
H_0: F_1(x) = f_1(x) \text{ and } F_2(x) = f_2(x),
\]
}
against the alternative hypothesis
{
\[
H_1: F_1(x) \neq f_1(x) \text{ or } F_2(x) \neq f_2(x).
\]
}

To evaluate this joint hypothesis, we first perform two hypothesis tests on each dataset pair to obtain their individual $p$-values. We then combine these $p$-values to assess the overall significance using Fisher's method. 
This two-step procedure is implemented using two distinct permutation-based metrics:

\begin{itemize}
    \item Energy Distance permutation test.
    \item Maximum Mean Discrepancy (MMD) permutation test.
\end{itemize}

\subsection{Energy Distance Permutation Test} 

The energy distance is a non-parametric statistic for quantifying the difference between two multivariate distributions. For simplicity in the mathematical formulation below, we use $X$ and $Y$ to denote a generic pair of observed and simulated datasets (e.g., $X_1$ and $Y_1$) of sizes $N$ and $M$. It compares the (log-transformed) pairwise distances between samples across datasets to the distances within each dataset. Intuitively, if the observed and simulated data are drawn from the same underlying distribution, the between-group and within-group distances should be statistically similar. In our case, a large energy distance indicates that the observed data are unlikely to originate from the simulation model. Specifically, if the observed discrepancy exceeds a threshold, we reject the null hypothesis that the two distributions are the same. The energy distance between real data and simulated data used here is adapted from \cite{zech2003newtestmultivariatetwosample}, with some minor modifications:
\begin{equation} \label{energy}
    D_{\text{energy}}(X, Y) = 2\,\frac{1}{NM}\sum_{i=1}^{N}\sum_{j=1}^{M}\log d(x_i,y_j) - \frac{1}{N(N-1)}\sum_{\substack{i,j=1 \\ i \neq j}}^{N}\log d(x_i,x_j) - \frac{1}{M(M-1)}\sum_{\substack{i,j=1 \\ i \neq j}}^{M}\log d(y_i,y_j).
\end{equation}
where $x_i$ is the sample from real data and $y_j$ is the sample from simulated data. $d(x, y)$ denotes the Mahalanobis distance, used to standardize the variables:
\[
d(x, y) = \sqrt{(x - y)^\top \hat\Sigma^{-1} (x - y)},
\]
with $\hat\Sigma$ being the estimated covariance of the simulated samples.

The test proceeds as follows:
\begin{enumerate}
    \item Sample $M$ data points $y_1, \ldots, y_M$ from the kernel density.
    \item Compute the observed $D^{\text{(obs)}}_{\text{energy}}$ between the $N$ observed data points $\{x_i\}_{i=1}^{N}$ and the $M$ sampled data points $\{y_j\}_{j=1}^{M}$ using Equation \ref{energy}.
    \item Pool all data (observed and sampled) together as
    \[
    \{z_k\}_{k=1}^{M+N} = \{x_1, \ldots, x_N, y_1, \ldots, y_M\}.
    \]
    \item Randomly permute the pooled set $P$ times. For the $p$th permutation, split $\{z_k\}$ into two groups of data $\{x^{(p)}_i\}_{i=1}^{N}$ and $\{y^{(p)}_j\}_{j=1}^{M}$ with the original sizes $N$ and $M$, and compute $D^{(p)}_{\text{energy}}$ using Equation \ref{energy}.
    \item The empirical $p$-value is computed as
    \[
    \text{$p$-value} = \frac{1}{P}\sum_{p=1}^{P} \mathbf{1}\Bigl\{ D^{(p)}_{\text{energy}} \geq D^{\text{(obs)}}_{\text{energy}} \Bigr\},
    \]
    where $\mathbf{1}\{\cdot\}$ is the indicator function.
\end{enumerate}

\subsection{Maximum Mean Discrepancy (MMD) Permutation Test}

The Maximum Mean Discrepancy (MMD) is a kernel-based method for comparing two distributions by measuring the distance between their mean embeddings in a reproducing kernel Hilbert space (RKHS), as proposed in \cite{JMLR:v13:gretton12a} for non-parametric two-sample testing. The MMD is defined as:
\[
\text{MMD}^2(X, Y) = \mathbb{E}_{X,X'}[k(X, X')] + \mathbb{E}_{Y,Y'}[k(Y, Y')] - 2\,\mathbb{E}_{X,Y}[k(X, Y)],
\]
where $X, X'$ and $Y, Y'$ are independent samples from distributions $P$ and $Q$, respectively, and $k(\cdot,\cdot)$ is a kernel function.

The empirical estimator of the MMD is given by:

\begin{equation} \label{MMD}
    \hat{\text{MMD}}^2(X, Y) = \frac{1}{N(N - 1)}\sum_{\substack{i,j=1 \\ i \neq j}}^{N} k(x_i, x_j) + \frac{1}{M(M - 1)}\sum_{\substack{i,j=1 \\ i \neq j}}^{M} k(y_i, y_j) - \frac{2}{NM}\sum_{i=1}^{N}\sum_{j=1}^{M} k(x_i, y_j).
\end{equation}

A common choice for the kernel is the Gaussian RBF kernel, which treats closer points as more similar:

\[
k(x, y) = \exp\left(-\frac{\|x - y\|_2^2}{2\sigma^2}\right),
\]

where $\sigma$ is the bandwidth parameter.

The permutation test for the MMD follows the same procedure as the Energy Distance test:
\begin{enumerate}
    \item Compute the observed $\hat{\text{MMD}}^2$ using Equation \ref{MMD}.
    \item Randomly permute the pooled data, split them into two groups of sizes $N$ and $M$, and compute $\hat{\text{MMD}}^2$ for each permutation.
    \item The empirical \( p \)-value is computed as
    \begin{equation}
        \text{\(p\)-value} = \frac{1}{P} \sum_{p=1}^{P} \mathbf{1} \left\{ \hat{\text{MMD}}^{2(p)} \geq \hat{\text{MMD}}^{2(\text{obs})} \right\},
    \end{equation}
    where \( \hat{\text{MMD}}^{2(\text{obs})} \) is the MMD computed on the original observed and simulated data, \( \hat{\text{MMD}}^{2(p)} \) is the MMD computed on the \( p \)th permutation, and \( \mathbf{1}\{\cdot\} \) is the indicator function.
\end{enumerate}

\section{Additional figures for simulated distributions in \texorpdfstring{$(\mu, \sigma)$}{musigma} space} \label{sec:appendix_B}

\begin{figure*}[h]
\centering
\includegraphics[scale = 0.5]{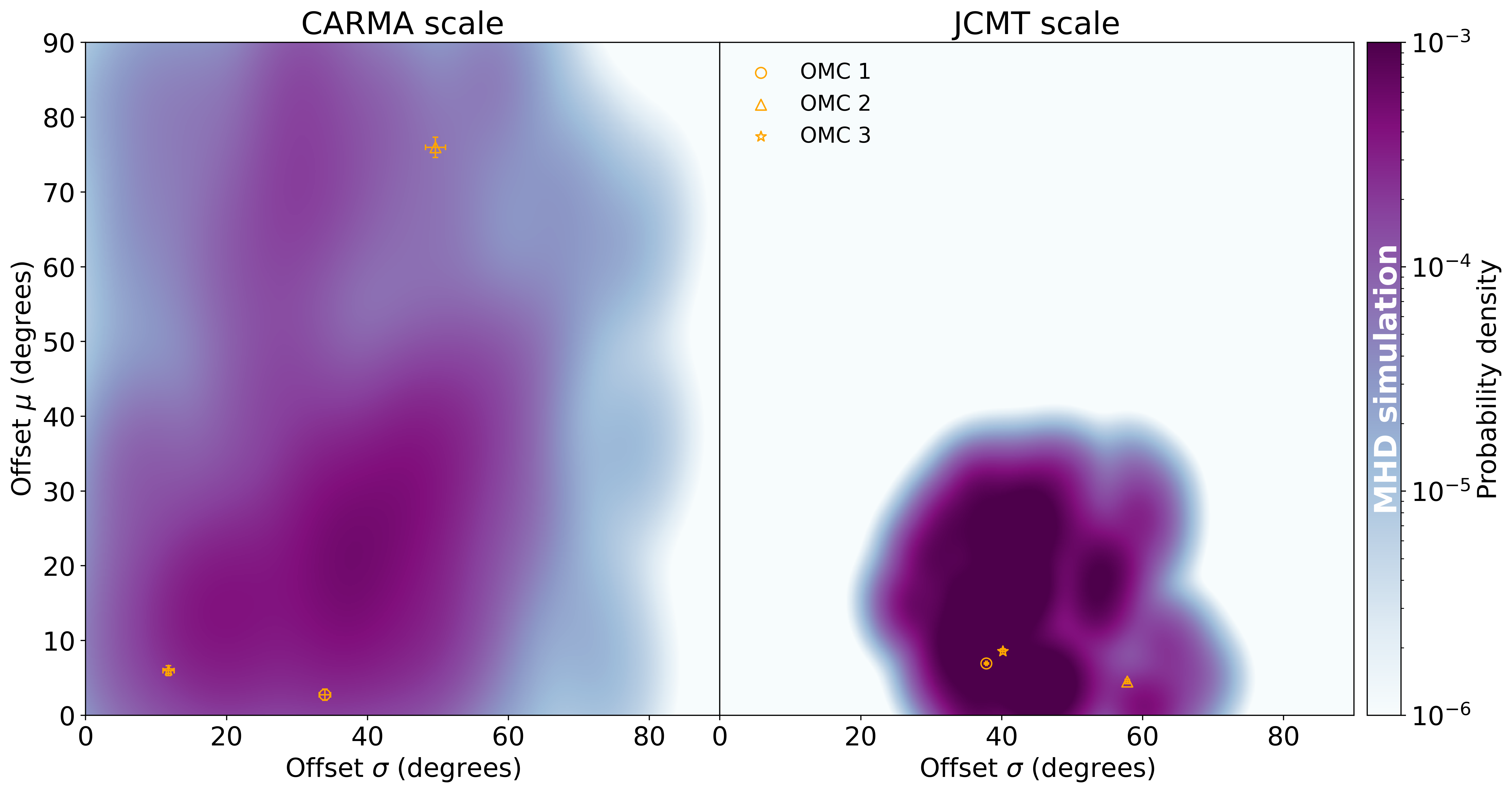}
\caption{An example of simulated distribution in $(\mu, \sigma)$ space created by aggregating many different samples in $\phi$ at $\theta = 35^\circ$. Made from simulation seed 1234.} \label{fig:simulated}
\end{figure*}

\begin{figure*}[ht]
\centering
\includegraphics[scale = 0.5]{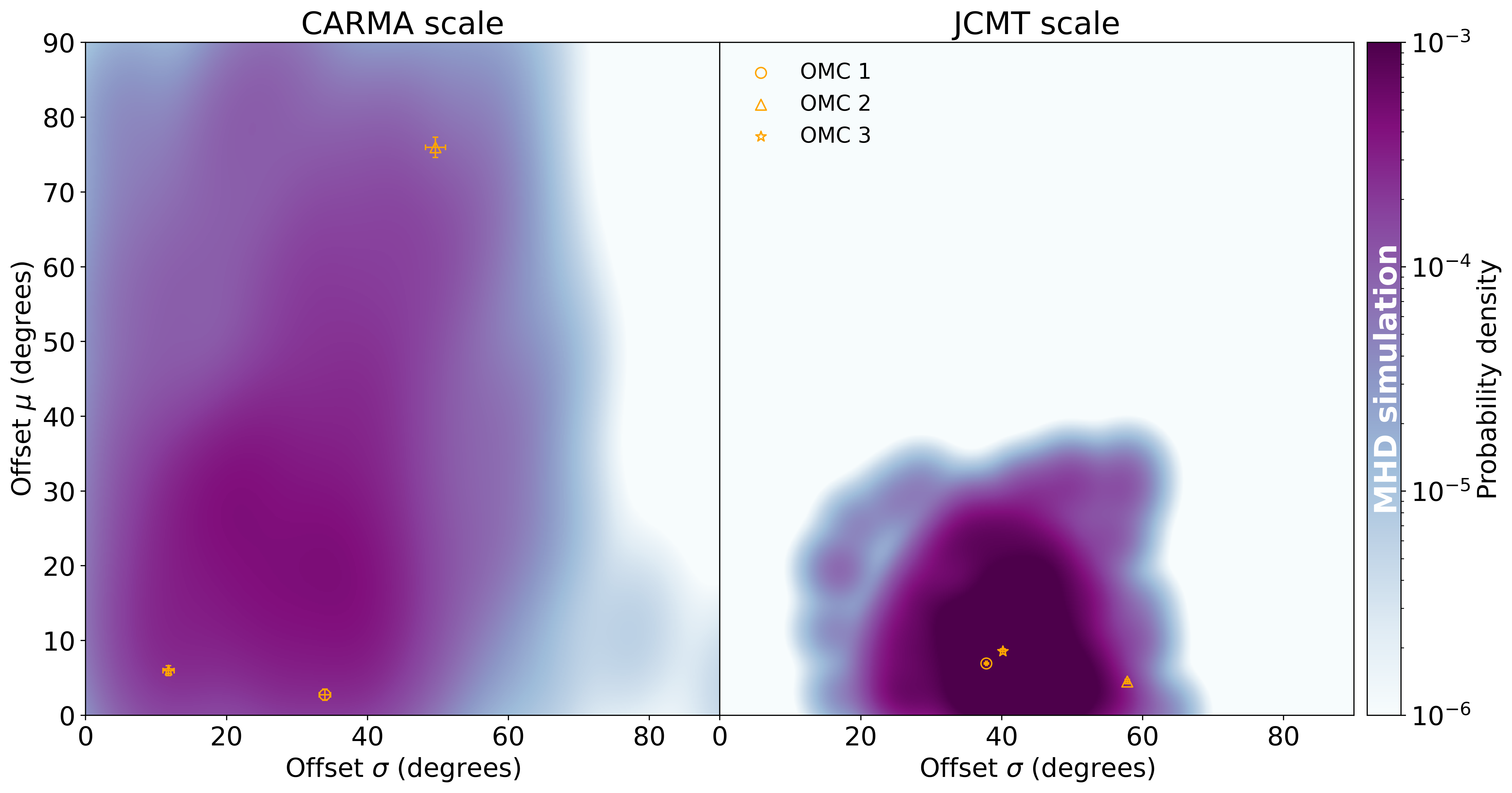}
\caption{Similar to Figure \ref{fig:simulated}, but made from simulation seed 7028 at $\theta = 35^\circ$.} \label{fig:fig11}
\end{figure*}

\begin{figure*}[h]
\centering
\includegraphics[scale = 0.5]{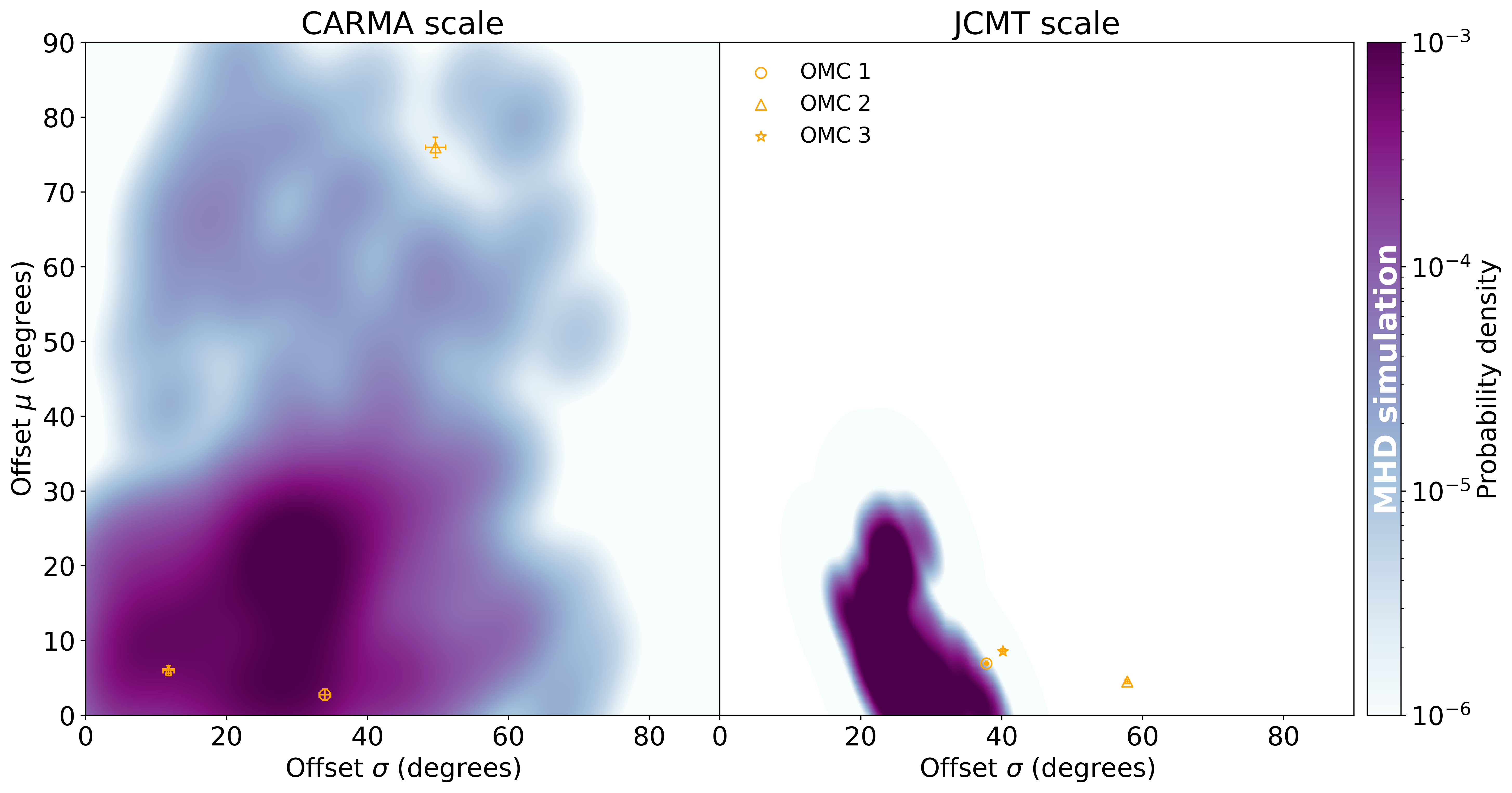}
\caption{Similar to Figure \ref{fig:simulated}, but made from simulation seed 1234 at $\theta = 80^\circ$.} \label{fig:fig12}
\end{figure*}

\begin{figure*}[h]
\centering
\includegraphics[scale = 0.5]{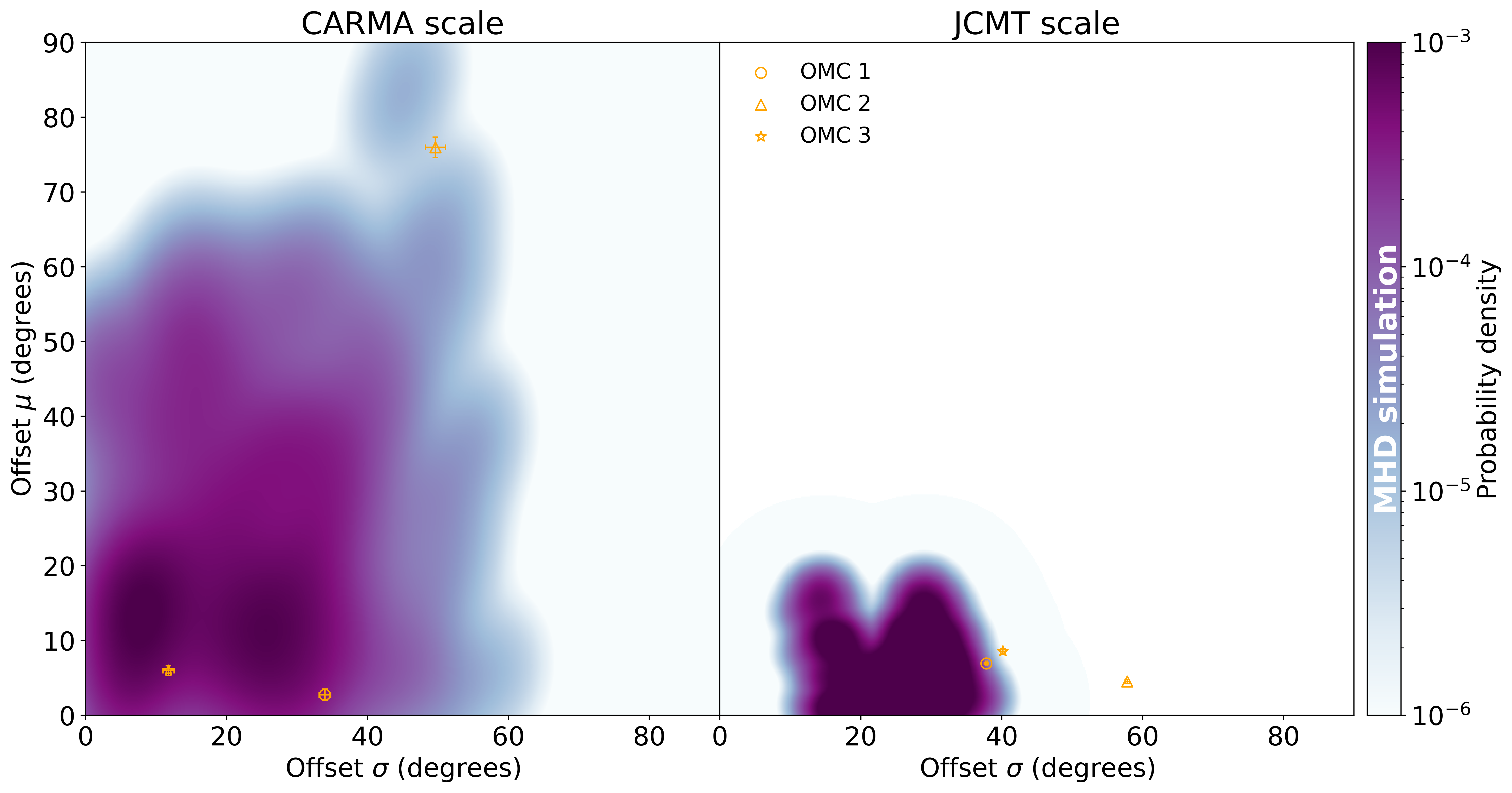}
\caption{Similar to Figure \ref{fig:simulated}, but made from simulation seed 7028 at $\theta = 80^\circ$.} 
\label{fig:fig13}
\end{figure*}

In this appendix, we provide additional figures to illustrate how the summary statistics $(\mu, \sigma)$ depend on the three dimensional orientation of the magnetic field relative to the line of sight. Each figure shows $(\mu, \sigma)$ distributions constructed by aggregating projections of simulation results over all azimuthal angles $\phi$ while holding the polar angle $\theta$ fixed. This procedure isolates the effect of the field -- line of sight inclination and removes degeneracy associated with azimuthal angles. Figures \ref{fig:simulated}-\ref{fig:fig13} compare results from two independent sub-Alfvénic MHD simulations. At a given $\theta$, the resulting $(\mu, \sigma)$ distributions are highly similar between simulations, demonstrating that these statistics are largely insensitive to the specific turbulence realization. In contrast, changing $\theta$ produces pronounced and systematic shifts in the $(\mu, \sigma)$ distributions. Similar to Figure \ref{fig:data_points}, the OMC observational data are overlaid in Figures \ref{fig:simulated}-\ref{fig:fig13}. In all cases, $\theta=35^\circ$ shows substantially better agreement with the observed  $(\mu, \sigma)$ values than $\theta=80^\circ$, and this conclusion holds independently of the chosen simulation.

\end{document}